\documentclass[a4paper,11pt]{article}
\pdfoutput=1 \usepackage{jheppub_2} 
\usepackage[T1]{fontenc} 
\usepackage[utf8]{inputenc}
\usepackage{color,subfigure}
\usepackage{graphicx, multirow,soul,url,amsmath,amsfonts,amssymb,mathrsfs,amsfonts}
\usepackage[all]{hypcap}
\usepackage{url}
\usepackage{bbm}
\usepackage{cancel}
\usepackage{subfigure}
\usepackage{slashed}
\usepackage[dvipsnames]{xcolor}
\usepackage{multicol, blindtext}
\usepackage[section]{placeins}
\usepackage[version=3]{mhchem}
\usepackage{caption}
\usepackage{lipsum}
\usepackage{hyperref}
\usepackage{float}
\usepackage{mattens}
\usepackage{mathtools}
\usepackage{footnote}
\usepackage{amsmath,amssymb,amsthm,amsxtra,overpic,bbm,bm,epsfig}
\usepackage{color,subfigure}
\usepackage[splitrule]{footmisc}
\usepackage{bbold}
\usepackage{amsmath}
\usepackage[normalem]{ulem}

\newcommand{\beq}{\begin{equation}}
\newcommand{\eeq}{\end{equation}}

\newcommand{\SU}{\,{\rm SU}}

\newcommand{\U}{\,{\rm U}}

\notoc 

\title{\boldmath The Higgs and Leptophobic Force at the LHC}

\author[a]{Pavel Fileviez P\'erez,}
\author[a]{Elliot Golias,}
\author[b]{Clara Murgui}
\author[a]{and Alexis D. Plascencia}
\affiliation[a]{Physics Department and Center for Education and Research in Cosmology and Astrophysics (CERCA), 
Case Western Reserve University, Cleveland, OH 44106, USA}
\affiliation[b]{Departamento de F\'isica Te\'orica, IFIC, Universitat de Valencia-CSIC, 
E-46071, Valencia, Spain}

\emailAdd{pxf112@case.edu}
\emailAdd{ebg23@case.edu}
\emailAdd{clara.murgui@ific.uv.es}
\emailAdd{alexis.plascencia@case.edu}

\abstract{The Higgs boson could provide the key to discover new physics at the Large Hadron Collider. We investigate novel decays of the Standard Model (SM) Higgs boson into leptophobic gauge bosons which can be light in agreement with all experimental constraints.
We study the associated production of the SM Higgs and the leptophobic gauge boson that could be crucial to test the existence of a leptophobic force.
Our results demonstrate that it is possible to have a simple gauge extension of the SM at the low scale, without assuming very small couplings and in agreement with all the experimental bounds
that can be probed at the LHC.}

\begin{document} 

\maketitle
\flushbottom

\newpage 

\section{Introduction}

The discovery of the Standard Model (SM) Higgs boson with a mass of 125 GeV at the Large Hadron Collider (LHC)~\cite{Aad:2012tfa,Chatrchyan:2012xdj} can be considered one of the most important discoveries in physics. 
We now understand how most of the elementary particles acquire mass through the Higgs mechanism and how the electroweak symmetry is spontaneously broken in nature.
Thanks to the great effort of the experimental collaborations at the LHC we know well the properties of the SM Higgs and there exist experimental constraints 
on its decays and production mechanisms, see for example Ref.~\cite{Tanabashi:2018oca} for a detailed discussion. 

The Higgs boson could open a door to a new physics sector since it can have new interactions that can provide information about a theory for physics beyond the Standard Model.
The LHC could discover new decays and/or production channels for the Higgs boson and combining different searches we could have access to new interactions and 
discover new particles with masses below the TeV scale. See Ref.~\cite{Cepeda:2019klc} for a report on future studies at the LHC. 

In this article, we investigate new possible decays and production mechanisms of the Higgs boson due to the existence of a new interaction with a leptophobic gauge boson. 
A leptophobic gauge boson is predicted in simple theories where baryon number is a local gauge symmetry~\cite{Pais:1973mi,Foot:1989ts,Carone:1995pu,FileviezPerez:2010gw} spontaneously broken at the low scale. See Refs.~\cite{FileviezPerez:2011pt,Duerr:2013dza,Perez:2014qfa} for realistic models predicting a leptophobic gauge boson and Ref.~\cite{Perez:2015rza} for a review.
In our studies we show that one can have a large branching ratio for the Higgs decays into two leptophobic gauge bosons if they are kinematically allowed. The leptophobic gauge boson 
can be light with mass below the electroweak scale in agreement with all experimental bounds and without assuming a very small gauge coupling.

When the new Higgs decays are highly suppressed or not allowed we investigate the associated Higgs-leptophobic gauge boson production mechanism at the LHC. 
We find that, using this production mechanism, one can obtain large number of events with multi-photons and two quarks that can be used to test the existence of a new interaction 
of the Higgs boson with this new gauge boson. As in the case of the Higgs decays, the production cross-sections can be generically large due to the fact that the leptophobic gauge boson can be light in agreement with all experimental bounds. 
The possible existence of a leptophobic gauge boson at the low scale tells us that a gauge theory where baryon number is a local symmetry~\cite{FileviezPerez:2010gw,FileviezPerez:2011pt,Duerr:2013dza,Perez:2014qfa} can describe physics below the TeV scale.

This article is organized as follows: In Section~\ref{sec:Leptophobic}, we review all current collider constraints on a leptophobic gauge boson and discuss the impact of these bounds on the predictions for production cross-sections at the LHC. In Section~\ref{sec:Decays}, we show the predictions for the new Higgs decay channels into two leptophobic gauge bosons 
taking into account all the experimental constraints.  In Section~\ref{sec:Associated}, we discuss the associated production channel proton-proton to the leptophobic gauge boson and the SM Higgs, $p p \to Z_B^* \to  Z_B h$, and investigate the different signatures at the LHC. We present our conclusions in Section~\ref{sec:Summary}. Appendices~\ref{app:decays} and~\ref{app:xsections} contain analytic results for all the processes considered in this work. In Appendix~\ref{app:kinmixing}, we discuss the bounds on the kinetic mixing between the $Z$ and the new gauge boson.

\section{Leptophobic Gauge Boson at the LHC }
\label{sec:Leptophobic}
In simple extensions of the SM where baryon number is a local symmetry~\cite{FileviezPerez:2010gw,FileviezPerez:2011pt,Duerr:2013dza,Perez:2014qfa} spontaneously 
broken one predicts the existence of a leptophobic gauge boson $Z_B$. For phenomenological studies of these models and dark matter see 
Refs.~\cite{Duerr:2014wra,Ohmer:2015lxa,Duerr:2017whl,FileviezPerez:2018jmr,FileviezPerez:2019jju}, while for a mechanism for baryogenesis in this scenario see Ref.~\cite{Carena:2019xrr}. The coupling between the SM quarks and $Z_B$ in our convention is given by
\begin{equation}
Z_B^\mu \bar{q} q : -i \frac{g_B}{3} \gamma^\mu.
\end{equation}
As we show in the following, the local baryon number can be broken at the low scale, even at energies below the electroweak scale.

The main strategy to search for a heavy $Z_B$ at the LHC is by looking for a dijet resonance. However, at low masses this search loses sensitivity due to the large QCD backgrounds. Nonetheless, recent experimental searches for a boosted leptophobic gauge boson  decaying into jets along with initial state radiation of a photon have been performed at CMS to place exclusion bounds down to a mass of 10 GeV for $Z_B$ \cite{Sirunyan:2019sgo}. This further motivates a study in the low mass region. 

\begin{figure}[b]
\centering
\includegraphics[width=0.95\linewidth]{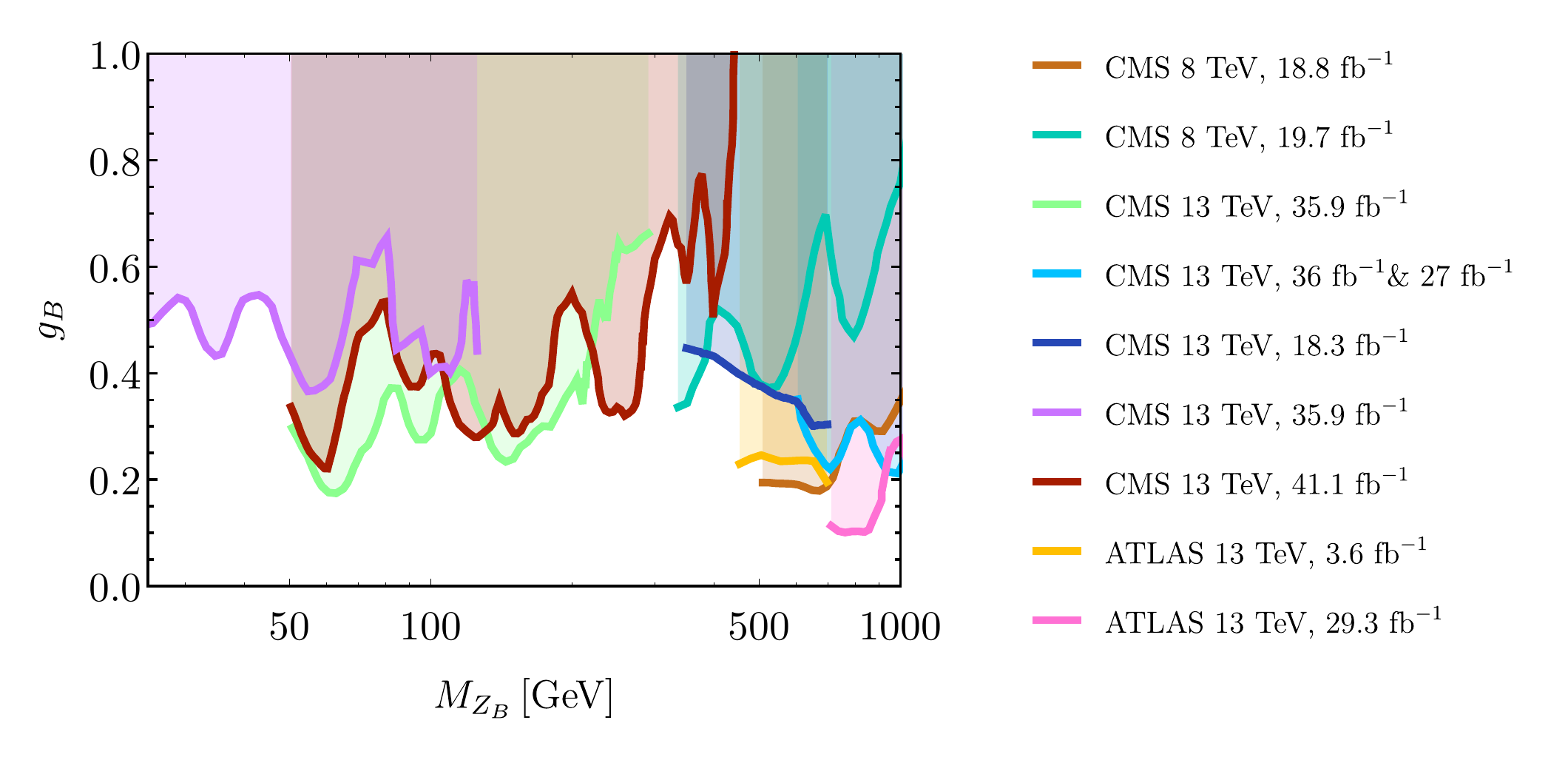}
\caption{Experimental bounds for the leptophobic gauge boson $Z_B$. Here, we use the CMS analyses (8 TeV and 18.8 $\text{fb}^{-1}$~\cite{Khachatryan:2016ecr}, 8 TeV and 19.7 $\text{fb}^{-1}$~\cite{Sirunyan:2018pas}, 13 TeV and 35.9 $\text{fb}^{-1}$~\cite{Sirunyan:2019sgo,Sirunyan:2019vxa} and 41.1 $\text{fb}^{-1}$~\cite{Sirunyan:2019vxa}, 13 TeV and 36 $\text{fb}^{-1}$ \& 27 $\text{fb}^{-1}$~\cite{Sirunyan:2018xlo}, 13 TeV and 18.3 $\text{fb}^{-1}$~\cite{Sirunyan:2019pnb}), and ATLAS results (13 TeV and 3.6 $\text{fb}^{-1}$ and 29.3 $\text{fb}^{-1}$~\cite{Aaboud:2018fzt}).} 
\label{ZBbounds}
\end{figure}

In Fig.~\ref{ZBbounds} we summarize the current collider bounds for the leptophobic gauge boson in the $g_B-M_{Z_B}$ plane. As this figure shows, there is a large region in the parameter space that remains unconstrained. Specifically, for a light $Z_B$ with mass between 25 and 50 GeV the gauge coupling can take relatively large values. For smaller couplings, i.e. $g_B\lesssim 0.1$, almost any value in the window $25 \, {\rm GeV} < M_{Z_B}<1$ TeV is allowed. Therefore, there is hope to
produce this gauge boson at the LHC with large cross-sections and study its properties.

\begin{figure}[h]
\centering
\includegraphics[width=0.52\linewidth]{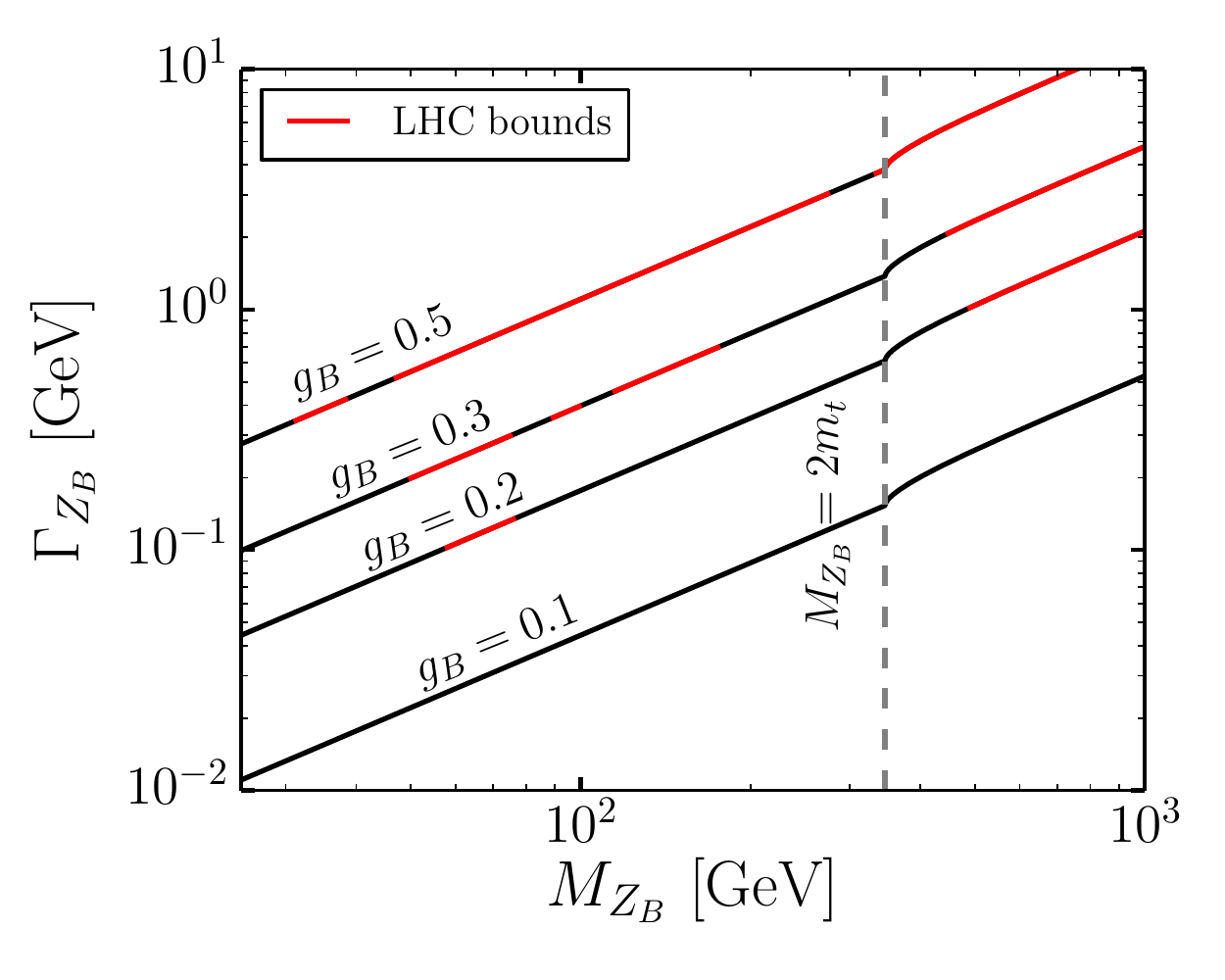}
\includegraphics[width=0.47\linewidth]{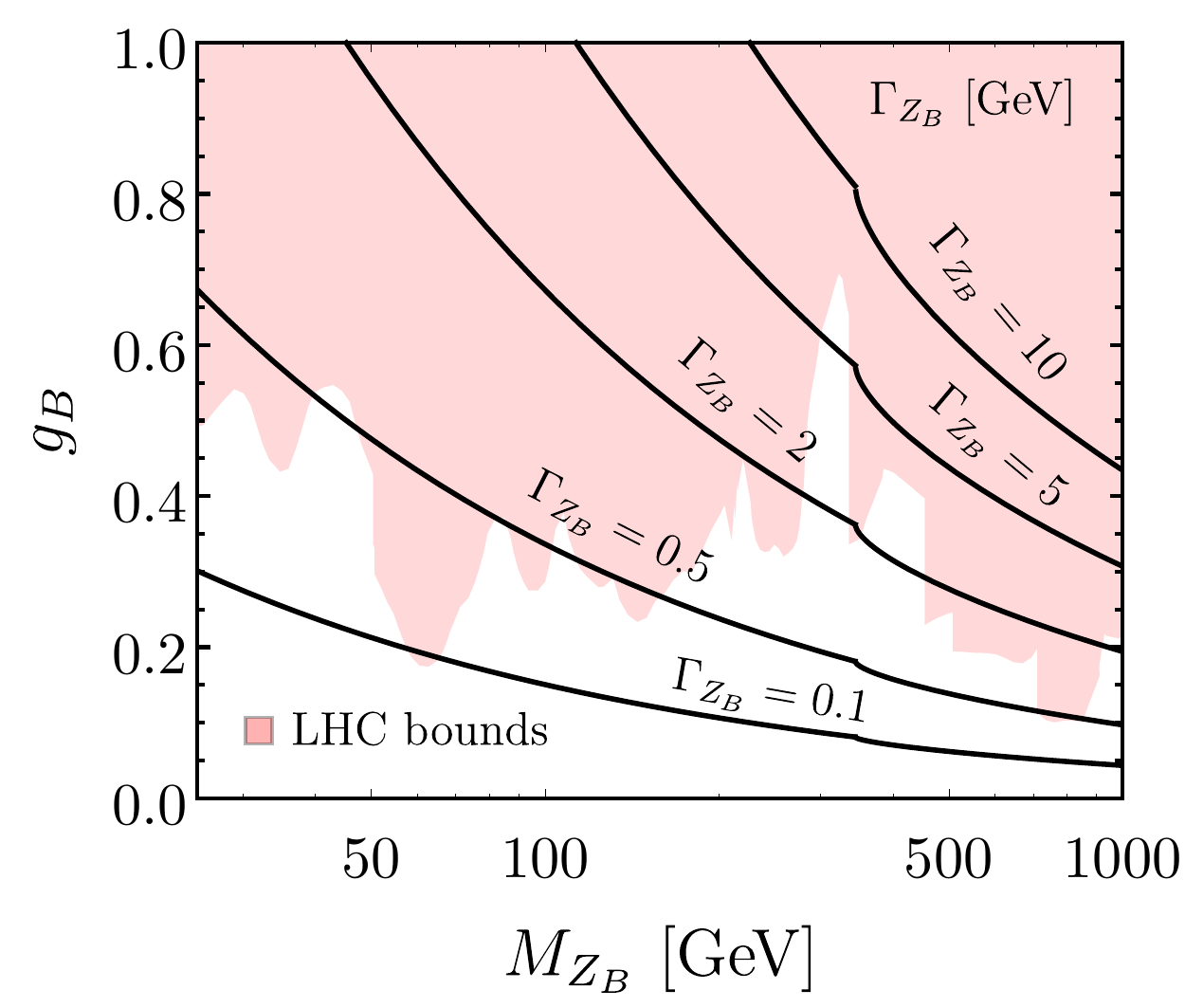}
\caption{\textit{Left panel:} Decay width of the $Z_B$ boson as a function of its mass. The regions highlighted in red are excluded by searches at the LHC. 
To the right of the dashed vertical line the decay channel $Z_B \to t \bar{t}$ is open. \textit{Right panel:} Contour lines for the decay width of the $Z_B$ boson in the $g_B$ vs $M_{Z_B}$ plane. } 
\label{ZBdecays}
\end{figure}

In the left panel in Fig.~\ref{ZBdecays} we show the decay width of $Z_B$ for different values of the gauge coupling $g_B$ as a function of its mass. 
In red we show the regions that are ruled out by the  collider bounds shown in Fig.~\ref{ZBbounds}. 
From this we can infer which are the allowed values for the decay width of the leptophobic gauge boson. Moreover, with this information of the decay width we can predict the different cross-sections relevant for different collider searches. In the right panel in Fig.~\ref{ZBdecays} we present contours of $\Gamma_{Z_B}$ in the $g_B$ vs $M_{Z_B}$ plane. The region shaded in red is excluded by collider searches of the $Z_B$ and we conclude that a $\Gamma_{Z_B}$ of order GeV is already mostly excluded.

\begin{figure}[h!]
\centering
\includegraphics[width=0.95\linewidth]{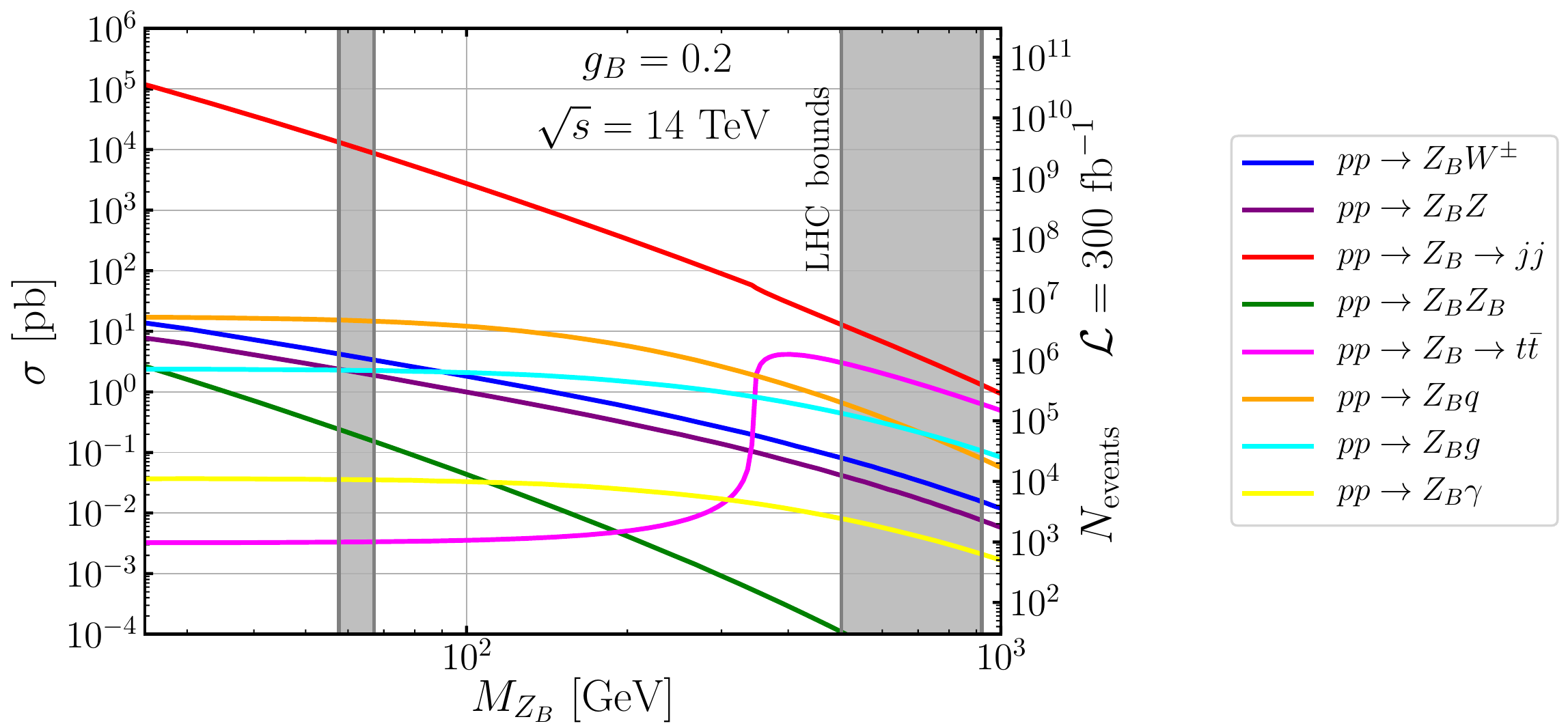}
\caption{Production cross-sections at the LHC for center-of-mass energy of 14 TeV in units of picobarns, we fixed $g_B\!=\!0.2$. On the right side of the vertical axis we show the expected number of events assuming $300$ fb$^{-1}$ for the integrated luminosity. The regions shaded in gray are excluded by LHC searches for the $Z_B$ boson.} 
\label{fig:xsections}
\end{figure}

In Fig.~\ref{fig:xsections} we present our results for the production cross-section for different channels that involve at least one $Z_B$, fixing the gauge coupling to $g_B=0.2$. These results correspond to the LHC with center-of-mass energy of 14 TeV and the number of events shown on the right vertical axis corresponds to an integrated luminosity of 300 fb$^{-1}$. The model has been implemented in \texttt{FeynRules 2.0}~\cite{Alloul:2013bka} and the cross-sections obtained using \texttt{MadGraph5aMC@NLO - v2.7.0}~\cite{Alwall:2014hca}, we cross-checked our results in a \texttt{Mathematica} notebook and the use of the MSTW2008~\cite{Martin:2009iq} set of parton distribution functions. In Appendices~\ref{app:decays} and~\ref{app:xsections} we provide analytic results for all the processes we have considered.

From Fig.~\ref{fig:xsections} one can see that the dijet cross-section dominates across the plot, and in the region $M_{Z_B}>2M_t$ the process $pp\to Z_B\to t\bar{t}$ can be large as well. The process $pp\to Z_B  q$ can be significant, since there is a large contribution from the parton distribution function of the gluon in the initial state. For the $pp\to Z_B\gamma, \, Z_B  q$ and  $Z_B  g $ channels we impose the following cuts on the rapidity and the transverse momentum:
$|\eta| < 2.5$, and $p_T> 150 \,\, {\rm GeV}$. These three channels are relevant for searches in the low mass regime.

\section{Exotic Decays of the SM-like Higgs}
\label{sec:Decays}
%
In extensions of the SM with a leptophobic gauge boson~\cite{FileviezPerez:2010gw,FileviezPerez:2011pt,Duerr:2013dza,Perez:2014qfa}, its mass generation comes from the vacuum expectation of a new Higgs boson with non-zero baryon number, and hence, the models have two Higgs scalars that can mix with each other. 
\begin{figure}[h]
\centering
\includegraphics[width=0.8\linewidth]{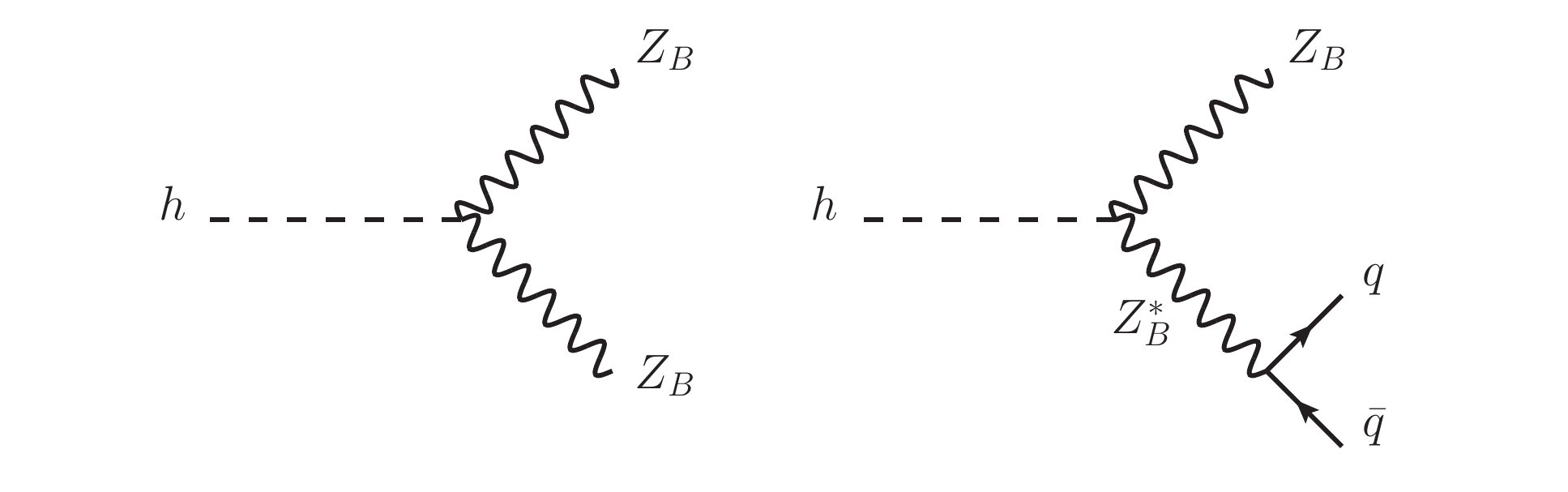}
\caption{Higgs decays into leptophobic gauge bosons.}
\label{hdecays}
\end{figure}
After spontaneous symmetry breaking, the SM-like Higgs will have the following coupling to the leptophobic $Z_B$ gauge boson
\begin{eqnarray}
\label{eq:hZBZB}
h Z_B^\mu Z_B^\nu: \,\, 2i \frac{M_{Z_B}^2}{v_B} g^{\mu \nu} \sin \theta_B, 
\end{eqnarray} 
where $\theta_B$ is the mixing angle in the scalar sector, $M_{Z_B}=Q_Bg_Bv_B$ and $Q_B$ is the baryon number of the second scalar.
Since the leptophobic gauge boson can be light, the SM-like Higgs can have the following decays
$$h \to Z_B Z_B, \ Z_B^* Z_B,$$
depending on the $Z_B$ mass, see Fig.~\ref{hdecays}. In order to calculate these decays one needs to know the coupling between the SM quarks
and the $Z_B$. Notice that the couplings between the SM-like Higgs and SM particles will scale by a factor $\cos \theta_B$.
With this information we can calculate the impact of these novel decays of the SM-like Higgs by computing the total Higgs decay width $\Gamma_h = \cos^2 \theta_B \Gamma_{\rm SM} + \Gamma_{\rm BSM}$, where in our case $\Gamma_{\rm BSM}$ corresponds to the decays into two leptophobic gauge bosons.

\begin{figure}[t]
\centering
\includegraphics[width=0.496\linewidth]{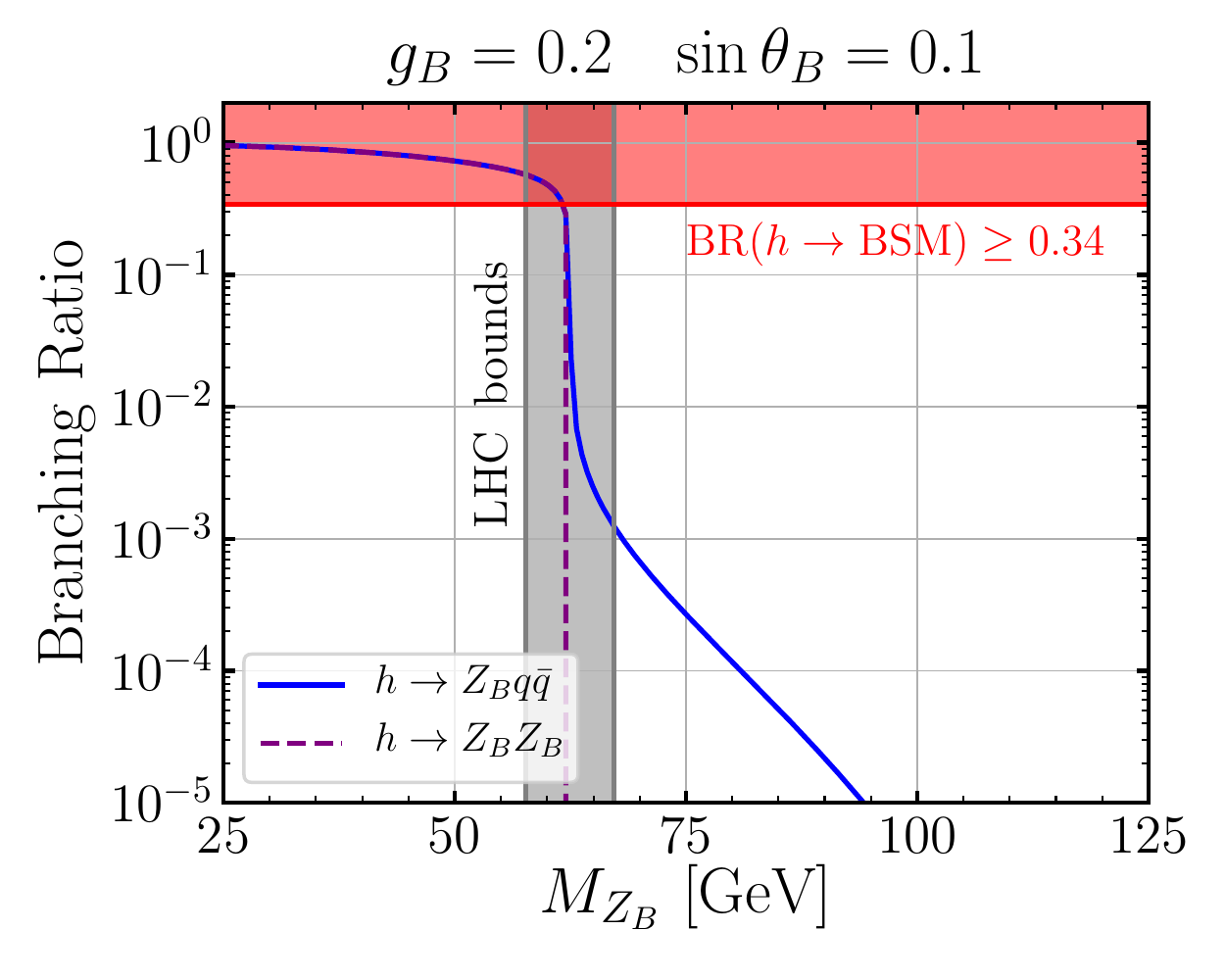}
\includegraphics[width=0.496\linewidth]{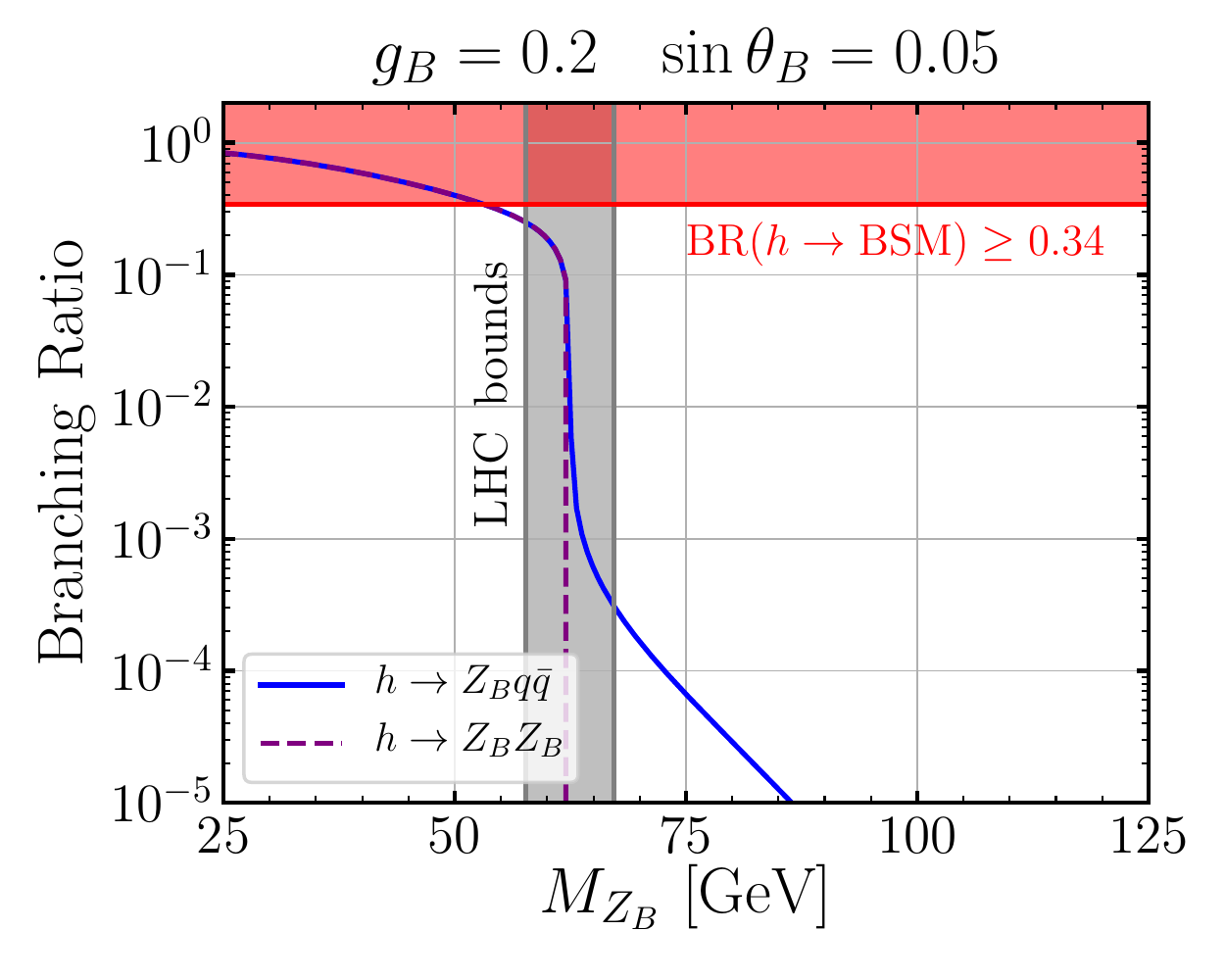}
\caption{Branching ratios for the channels $h \to Z_B Z_B$ and $h \to Z_B Z_B^* \to Z_B q \bar{q}$. The left (right) panel corresponds to $g_B=0.2$ and $\sin \theta_B=0.1$ ($\sin \theta_B=0.05$). The region shaded in red shows the exclusion bounds from the constraint on the SM-like Higgs branching ratio BR($h \to $ BSM) < 0.34. The region shaded in gray corresponds to the exclusion bounds from direct searches for the $Z_B$ boson at the LHC.} 
\label{fig:BR}
\end{figure}
\begin{figure}[t]
\centering
\includegraphics[width=0.496\linewidth]{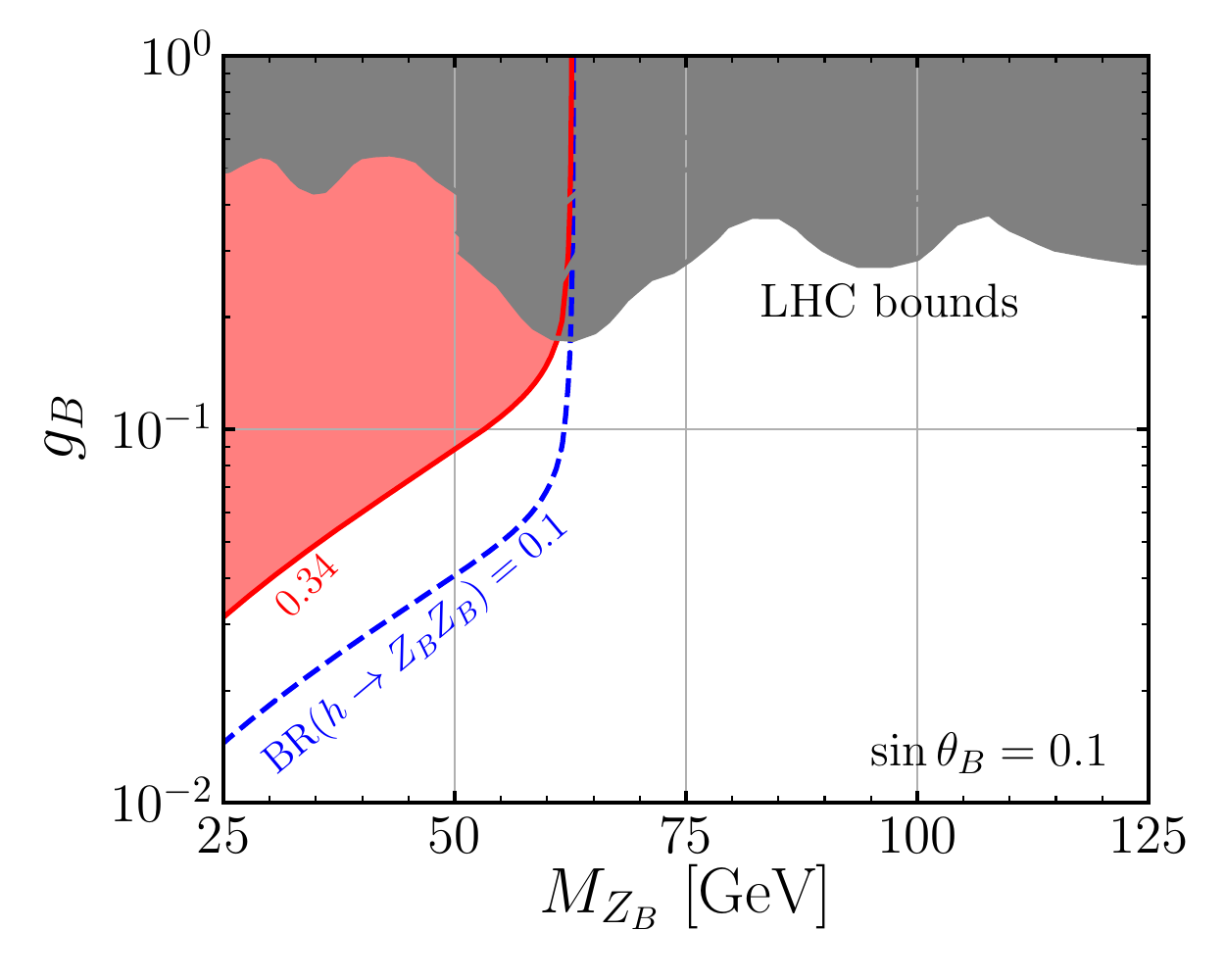}
\includegraphics[width=0.496\linewidth]{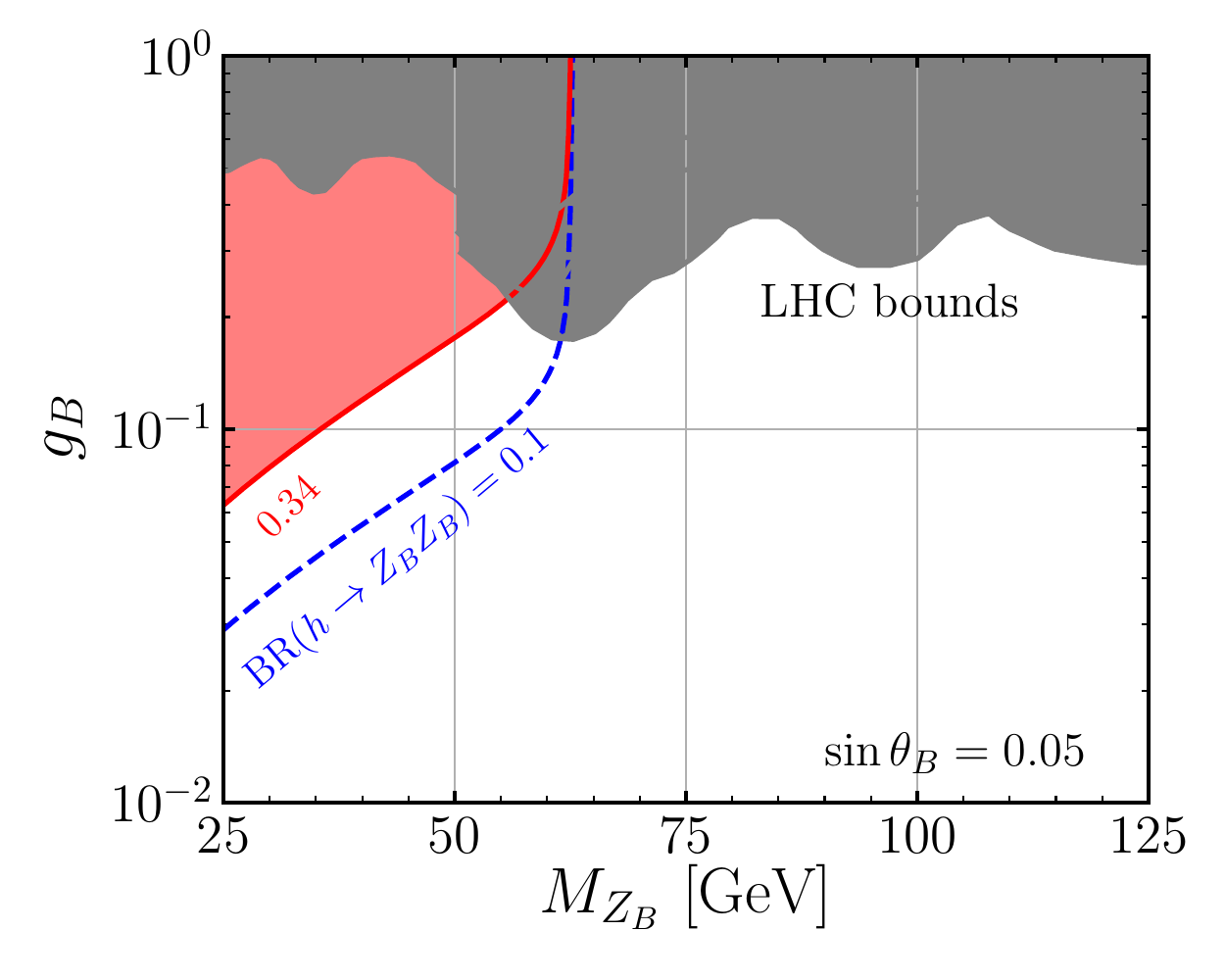}
\caption{Exclusion bounds in the $g_B$ vs $M_{Z_B}$ plane. The region shaded in red shows the exclusion bounds from the constraint on the SM-like Higgs branching ratio BR($h \to Z_B Z_B$) < 0.34, while the blue dashed line corresponds to BR($h \to Z_B Z_B)=0.1$. The region shaded in gray is excluded by searches for the $Z_B$ at the LHC. The left (right) panel corresponds to $\sin \theta_B=0.1$ ($\sin \theta_B=0.05$).} 
\label{fig:gBBR}
\end{figure}

Collider searches of a new scalar mixing with the SM Higgs combined with measurements of the SM
Higgs properties provide constraints on the mixing angle. In our study we take the bound $\sin \theta_B \leq 0.3$ \cite{Ilnicka:2018def}. Current LHC measurements of the properties of the SM-like Higgs boson give the following constraint on its branching ratio into BSM particles \cite{Khachatryan:2016vau}
\beq
{\rm BR}(h\to {\rm BSM}) < 0.34 \hspace{0.5cm} {\rm at} \,\,  95\% \,\, {\rm CL},
\eeq
which is obtained assuming the production of the Higgs in the SM. Therefore, we scale the bound by the ratio between the production cross-section for the Higgs in the SM with the one in this model, which is given by ${\rm BR}(h\to {\rm BSM}) < 0.34 \times \left( \sigma_h^{\rm SM}/\sigma_h \right) = 0.34/\cos^2\theta_B$.

\begin{figure}[t]
\centering
\includegraphics[width=0.9\linewidth]{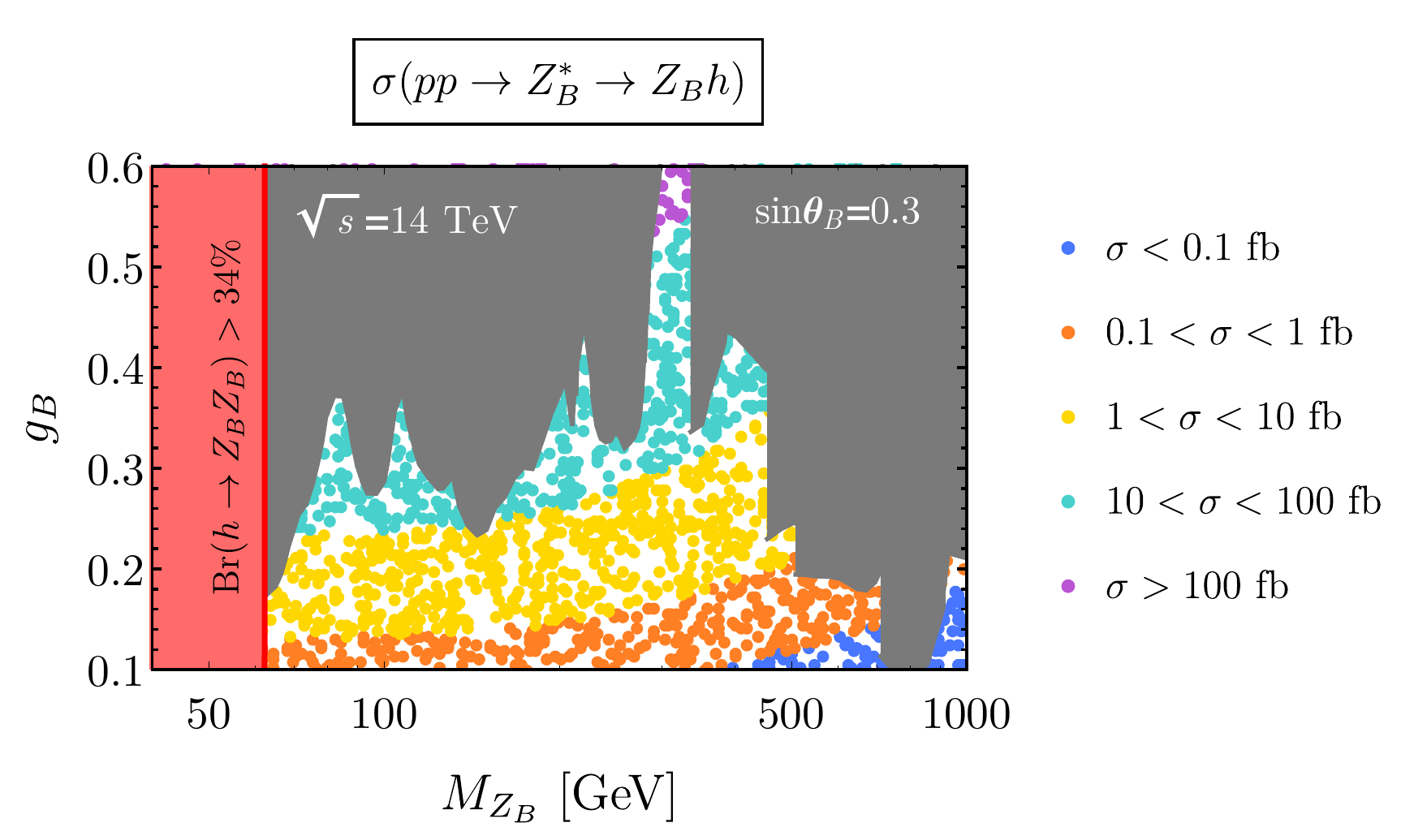}
\caption{Predictions for the associated production cross-section $p \ p \to Z_B^* \to Z_B \ h$ at the LHC with center-of-mass energy of 14 TeV. The gray region is excluded by the LHC bounds, while the red region is excluded 
by the bound on the branching ratio of the new Higgs decays. The scalar mixing angle is fixed to $\sin \theta_B=0.3$ for this plot.}
\label{sigmaZBh}
\end{figure}

We have computed the two-body and three-body decays and provide analytic expressions in Appendix~\ref{app:decays}.
In Fig.~\ref{fig:BR} we present our results for the branching ratios for the decay channels $h \to Z_B Z_B$ and $h \to Z_B q \bar{q}$ of the SM Higgs. The latter includes both, the on-shell and the off-shell contribution from the $Z_B$. In the region with $M_{Z_B}\leq M_h/2 \approx 62.5$ GeV the channel $h \to Z_B Z_B$  becomes the dominant decay channel and the Higgs decay width can become of order GeV. In this region the bound on BR($h \to $ BSM) < 0.34 gives a strong constraint shown by the area shaded in red. The gray band in this figure corresponds to the exclusion bounds from direct searches for the $Z_B$ boson at the LHC discussed in Section~\ref{sec:Leptophobic}.

On the other hand, when $M_{Z_B}\geq M_h/2$ the two-body decay is kinematically closed and the three-body decay gives a much smaller contribution to the Higgs width. In this regime, experiments can search for the associated Higgs $Z_B$ production to probe the existence of these interactions, as we discuss in the following section.

The experimental bound on the branching ratio of Higgs decays to BSM particles can be translated to the $g_B$ vs $M_{Z_B}$ plane. Nevertheless, we note that this bound also depends on the scalar mixing. In Fig.~\ref{fig:gBBR} we present our results for two different mixing angles. For $\sin \theta_B=0.1$ this constraint is strong in the region $M_{Z_B}\leq M_h/2$ and excludes $g_B \gtrsim 0.03$ for $M_{Z_B} =  25$ GeV. In order to relax this bound one needs to go to very small mixing angles,  $\sin \theta_B<0.05$, as shown in the right panel. It is important to emphasize that the SM-like Higgs can have a large branching ratio into two leptophobic gauge bosons in agreement with all current experimental bounds.

\section{Higgs-Leptophobic Gauge Boson Associated Production}
\label{sec:Associated}

\begin{figure}[b]
\centering
\includegraphics[width=0.35\linewidth]{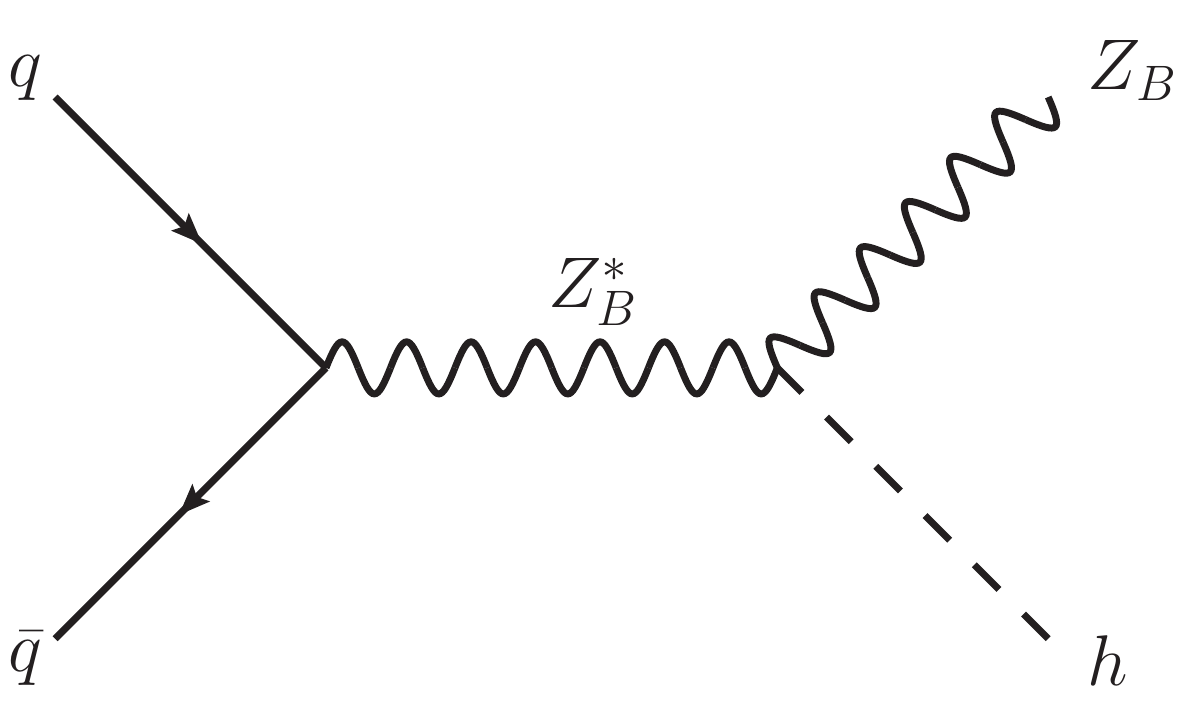}
\caption{Associated $Z_B - h$ production channel.}
\label{hZBproduction}
\end{figure}

In the previous section we discussed the possible new Higgs decays due to the existence of a leptophobic gauge boson. 
In the scenarios where these Higgs decays are not allowed or highly suppressed, one can study the associated production 
$$p \ p \to Z_B^* \to Z_B \ h, $$
to test the existence of the new $h-Z_B-Z_B$ interaction. See Fig.~\ref{hZBproduction} for the relevant Feynman graph.
The production cross-section for this process is given by Eq.~\eqref{ZBhproduction}.  In Fig.~\ref{sigmaZBh} we show the numerical 
predictions for the associated production $p \ p \to Z_B^* \to Z_B \ h$ in the $g_B-M_{Z_B}$ plane, in the maximal mixing scenario where $\sin \theta_B=0.3$ and with center-of-mass energy of $\sqrt{s}=14$ TeV.
The region shaded in red is excluded by the experimental bound on the branching ratio of the SM Higgs into BSM particles discussed in the previous section.
The different colored dotted regions correspond to the predictions in different ranges: $\sigma < 0.1 \ \rm{fb}$ (blue dots), $0.1  \ \rm{fb} < \sigma < 1 \ \rm{fb}$ (orange dots), 
$1  \ \rm{fb} < \sigma < 10 \ \rm{fb}$ (yellow dots), $10  \ \rm{fb} < \sigma < 100 \ \rm{fb}$ (cyan dots), and $\sigma > 100 \ \rm{fb}$ (purple dots). 
The production cross-section can easily be in the tens of femtobarns which is not too far from the $pp\to Z h$ cross-section of 990.33 fb in the SM~\cite{deFlorian:2016spz}.
The region shaded in gray is excluded by the collider bounds discussed in Section~\ref{sec:Leptophobic}. The associated cross-section is proportional to 
$\sin^2 \theta_B$. Therefore, although in the above figure we show only the predictions for $\sin \theta_B=0.3$, one can easily find the predictions  values for other mixing angles.
It is important to mention that the associated production can be significant due to the fact that the gauge coupling can be large and the mass of the leptophobic 
gauge boson can be below the electroweak scale.

\newpage
\begin{figure}[ht]
\centering
\includegraphics[width=0.45\linewidth]{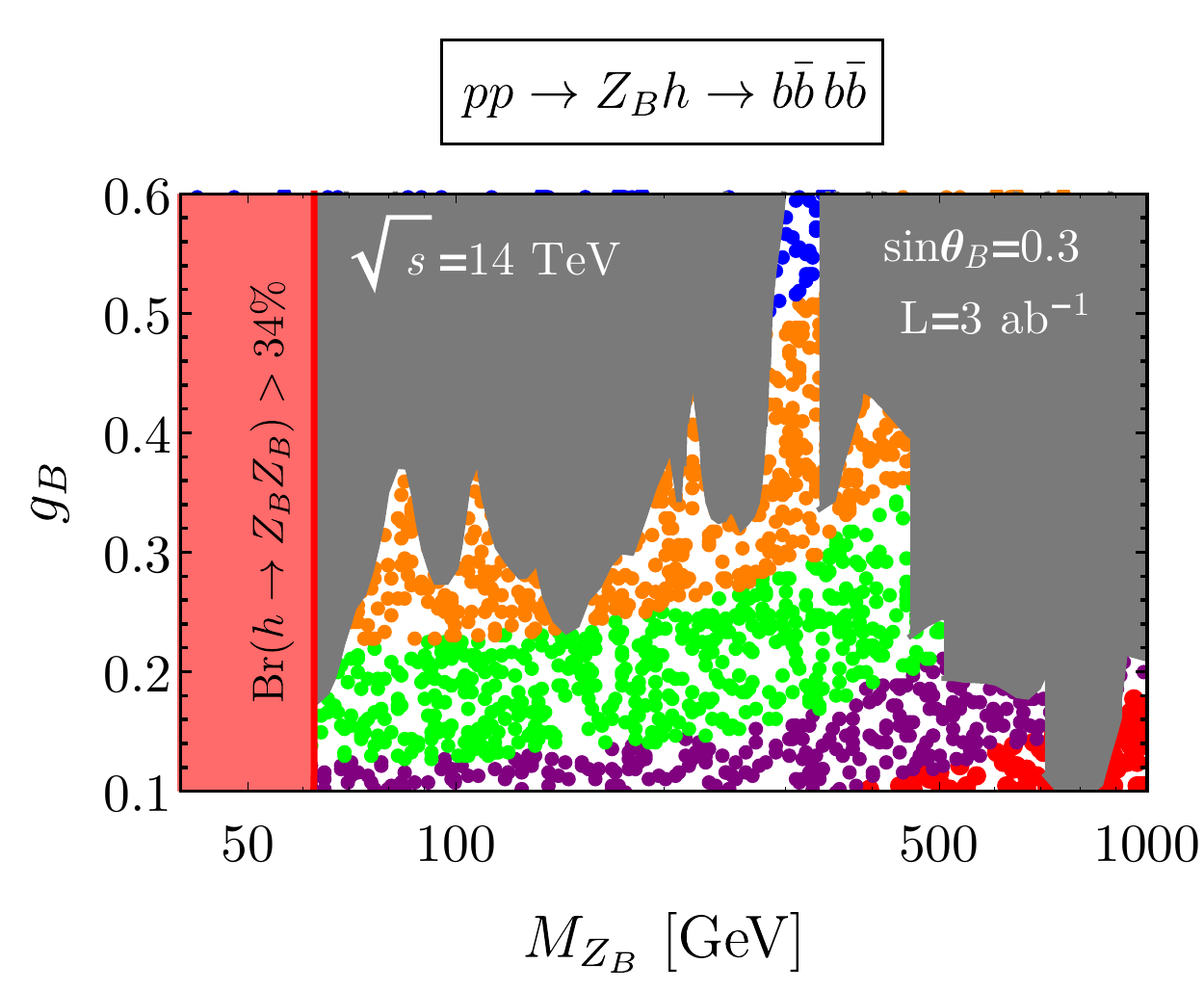}
\includegraphics[width=0.45\linewidth]{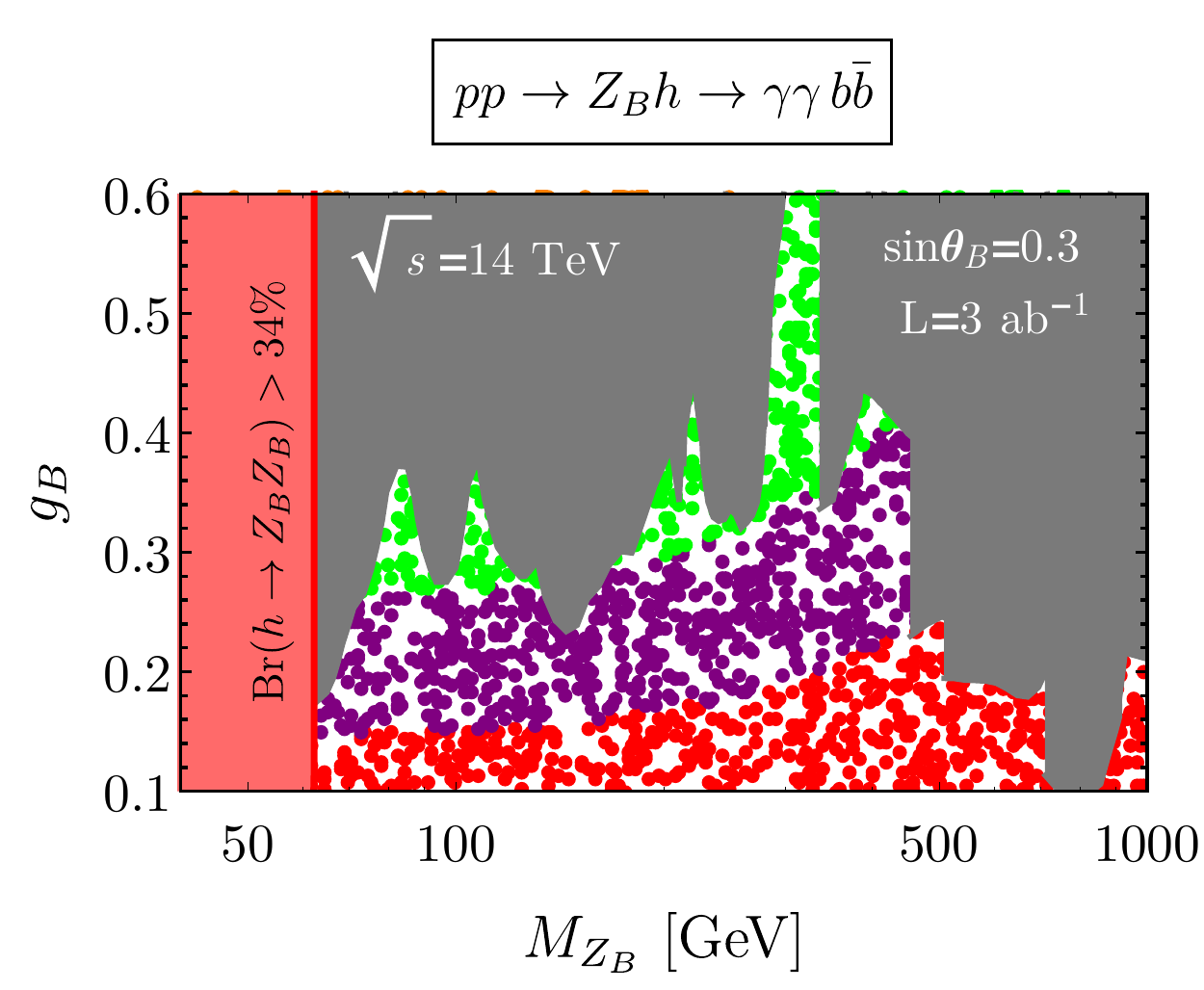}
\includegraphics[width=0.45\linewidth]{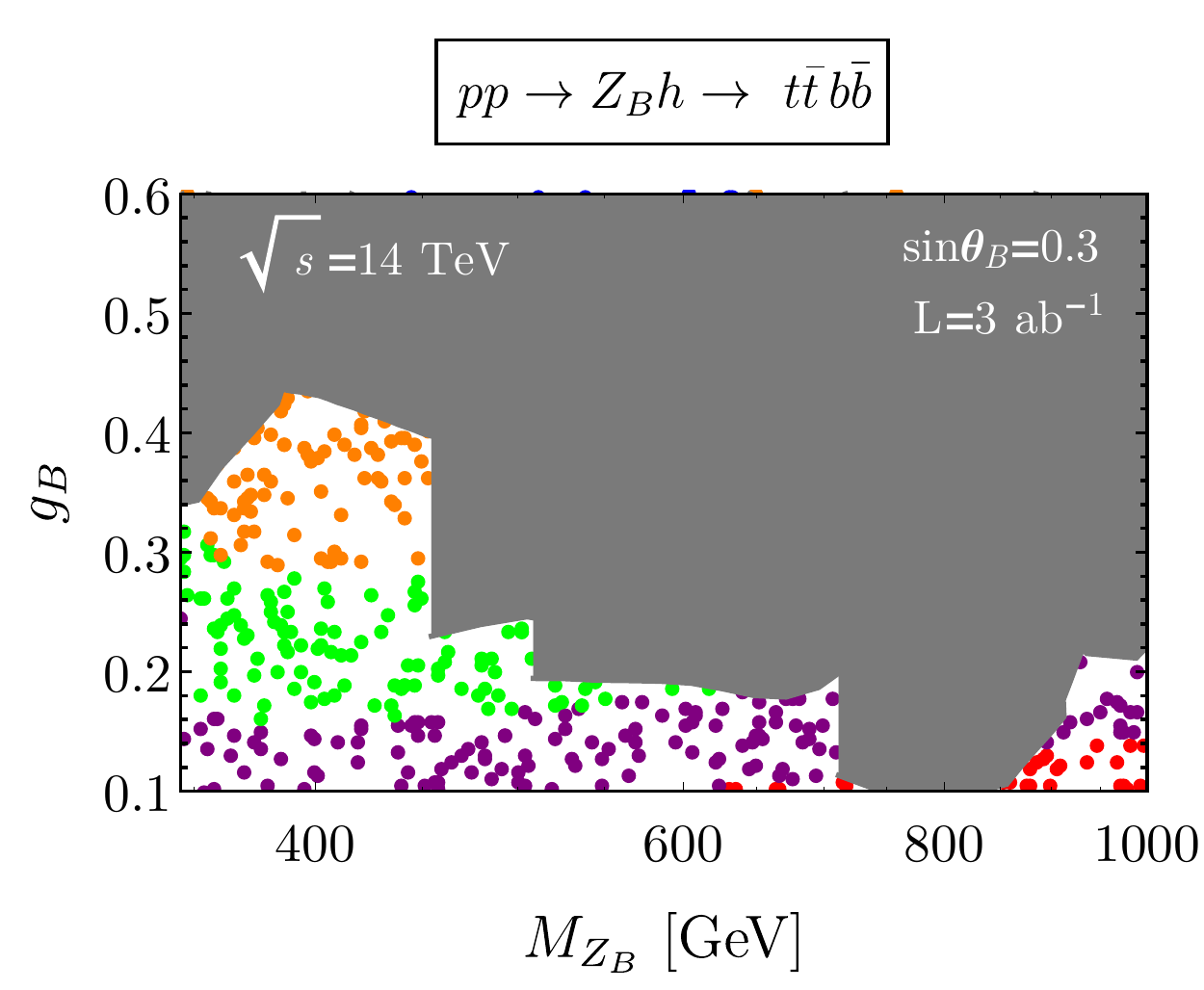}
\includegraphics[width=0.45\linewidth]{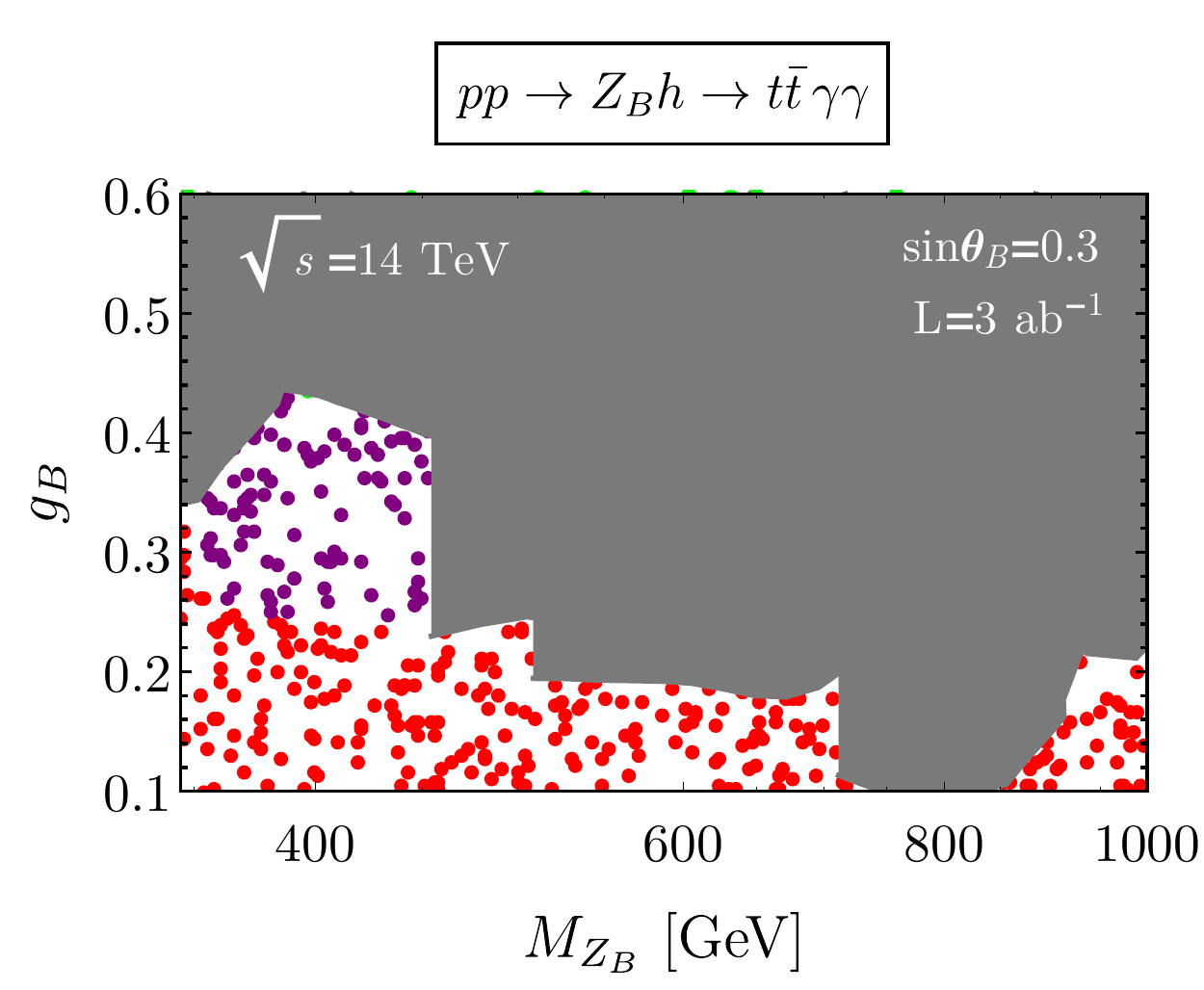}
\includegraphics[width=0.45\linewidth]{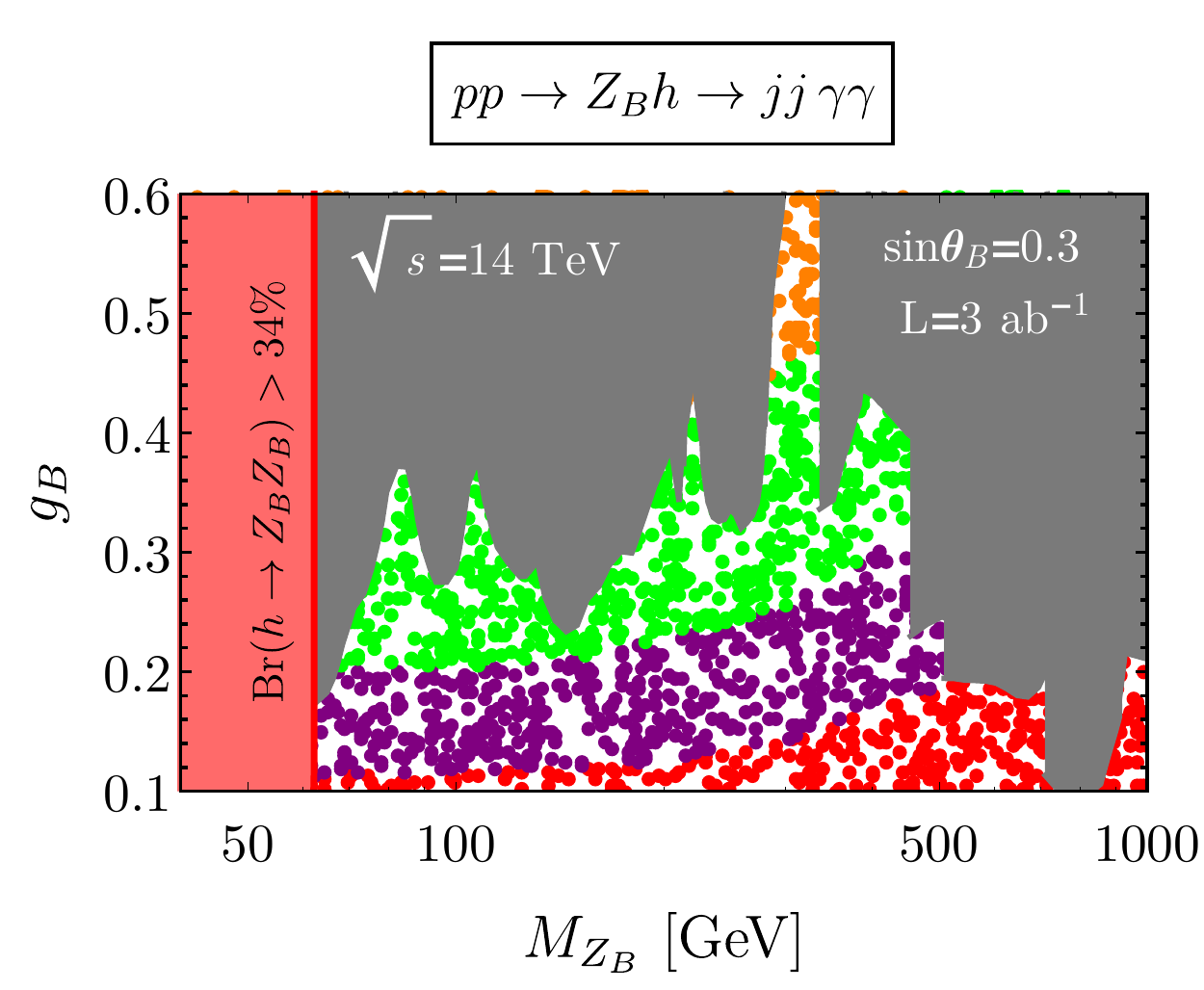}
\includegraphics[width=0.45\linewidth]{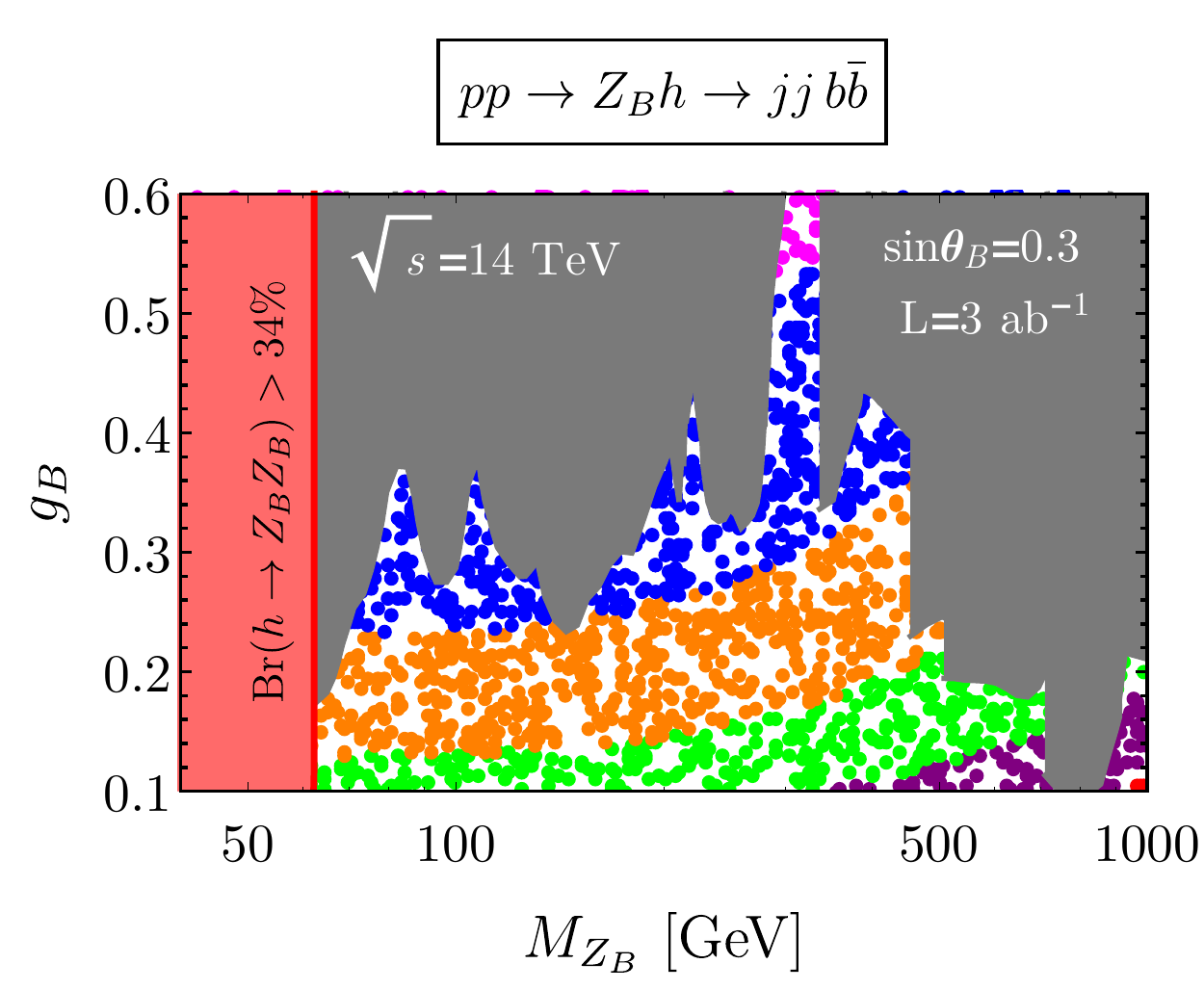}
\includegraphics[width=0.7\linewidth]{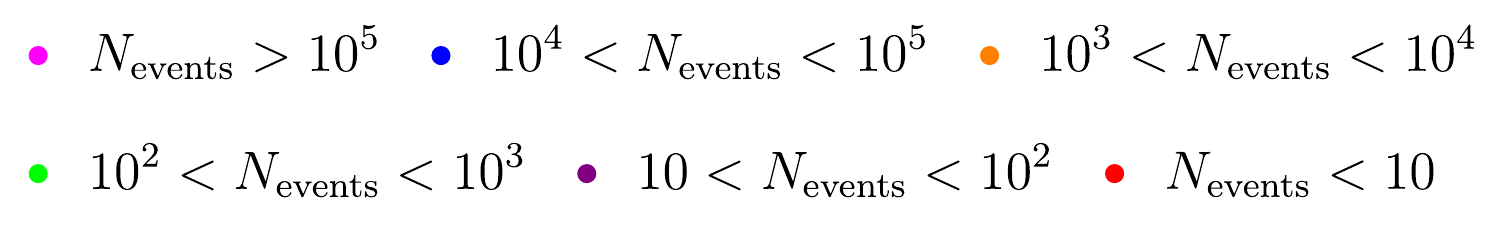}
\caption{Predictions for the number of events at the LHC with center-of-mass energy of 14 TeV assuming that the integrated luminosity is ${\cal{L}}=3000\  \rm{fb}^{-1}$ and using the maximal allowed value for the mixing 
angle $\sin \theta_B=0.3$. We show the number of events for the most relevant channels: $\gamma \gamma \ t \bar{t}$, $\gamma \gamma \ b \bar{b}$, 
$\gamma \gamma \ j j$, $b \bar{b} b \bar{b}$,  $b \bar{b} t \bar{t}$, and $b \bar{b} jj$. The gray region is excluded by the LHC bounds, while the red region is excluded 
by the bound on the branching ratio of the new Higgs decays.}
\label{Nevents}
\end{figure}
\newpage

Knowing the possible $h$ and $Z_B$ decays we can show the predictions for the number of events at the LHC for the following channels:
$$\gamma \gamma \ t \bar{t}, \gamma \gamma \ b \bar{b}, \  \gamma \gamma \ j j, \ b \bar{b} b \bar{b}, \ b \bar{b} t \bar{t}, \  {\rm{and}} \ b \bar{b} jj.$$
The number of events for each of these channels is given by
\begin{equation}
N_{\rm events}(x\bar{x} y \bar{y})= {\mathcal{L}}  \times \sigma (p \ p \to Z_B^* \to Z_B \ h)  \times {\rm BR}(h\to x\bar{x}) \times { \rm BR}(Z_B\to y \bar{y}).
\end{equation}
In Fig.~\ref{Nevents}  we show the predictions for the expected number of events assuming that the integrated luminosity is ${\cal{L}}=3000 \  \rm{fb}^{-1}$ as planned for the High-Luminosity LHC~\cite{ApollinariG.:2017ojx},
and using the maximal allowed value for the mixing angle $\sin \theta_B=0.3$. The gray regions in Fig.~\ref{Nevents} are excluded by the collider bounds 
discussed in Section~\ref{sec:Leptophobic}, while the regions in red are excluded by the experimental bound on the branching ratio of SM Higgs exotic decays.

The $Zh$ associated production has been measured at ATLAS~\cite{Aaboud:2018zhk} and CMS~\cite{Sirunyan:2017elk}, and a similar technique can be used to make the reconstruction of the processes in Fig.~\ref{Nevents}. However, in our case the $Z_B$ decays only to quarks and then the QCD background is more challenging. For example, the largest number of events is for the channel: $pp \to Z_B h \to jj b \bar{b}$. In this case two $b$ jets should have an invariant mass around the Higgs mass of 125 GeV. Furthermore, the large $p_T$ of the Higgs or the gauge boson can help discriminate the signal with respect to the background~\cite{Butterworth:2008sd}. A dedicated analysis for these signatures is required but it is beyond the scope of this article.

\section{Summary}
\label{sec:Summary}
The SM Higgs boson can open a doorway to new physics and there is a chance to discover a new sector from the existence of new interactions with the Higgs. In this article, we investigated the possibility that the Higgs can have a new interaction with a leptophobic gauge boson. In this scenario, Higgs decays can have a large branching ratio into two leptophobic gauge bosons if they are kinematically allowed. The leptophobic gauge boson 
can be very light, with mass below the electroweak scale, in agreement with all experimental bounds and without assuming a very small gauge coupling.

In the case where the new Higgs decays are highly suppressed or not allowed, we investigated the associated production of the Higgs and the leptophobic gauge boson at the LHC. 
We showed that this channel can lead to a large number of events with multi-photons and two quarks, which can be used to probe the existence of the interaction of the Higgs with the new gauge boson.
As in the case of the exotic Higgs decays, the production cross-sections can be generically large due to the fact that the leptophobic gauge boson can be light in agreement with all experimental bounds. It is relevant to mention that the possible existence of a leptophobic gauge boson at the low 
scale tells us that it is possible to have a simple gauge theory where baryon number is a local gauge symmetry~\cite{FileviezPerez:2010gw,FileviezPerez:2011pt,Duerr:2013dza,Perez:2014qfa} describing physics below the TeV scale.

\vspace{1cm}

{\small{{\textit{Acknowledgments}}: The work of P.F.P. has
been supported by the U.S. Department of Energy, Office of Science, Office of High Energy
Physics, under Award Number DE-SC0020443. The work of C.M. has been supported in
part by Grants No. FPA2014-53631-C2-1-P, No. FPA2017-84445-P, and No. SEV-2014-
0398 (AEI/ERDF, EU), and by La Caixa-Severo Ochoa scholarship.}}

\appendix

\section{Decays Widths}
\label{app:decays}

\begin{itemize}

\item Leptophobic Gauge Boson:

\begin{figure}[h]
\centering
\includegraphics[width=0.35\linewidth]{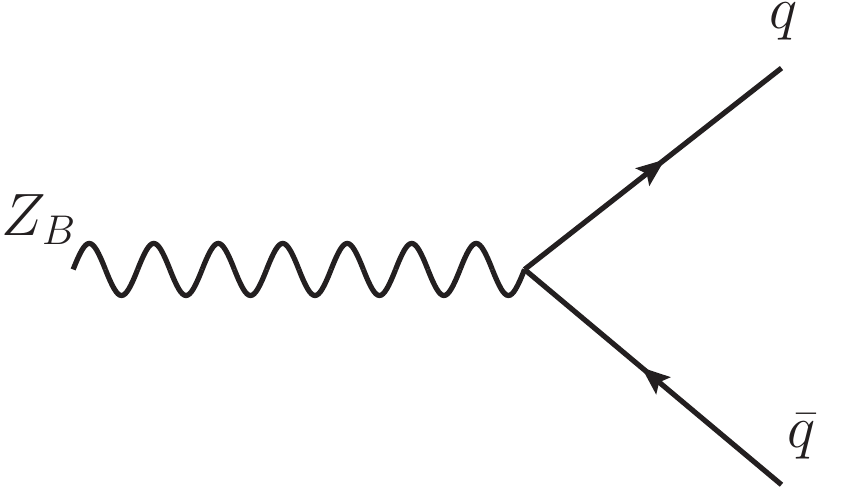}
\caption{Leptophobic gauge boson decay.}
\end{figure}

The partial decay width of the leptophobic gauge boson $Z_B$ with mass $M_{Z_B}$ is given by
\begin{align}
  \Gamma ( Z_B \rightarrow \bar{q} q ) &= \frac{g_B^2 M_{Z_B}}{36 \pi}  \sqrt{1 - \frac{4 M_q^2}{M_{Z_B}^2} } \left(1 + \frac{2 M_q^2}{M_{Z_B}^2} \right),
 \end{align}
where $M_q$ is the mass of a given quark. 

\item New Higgs Decays:

The width for the new two-body decays, $h \to Z_B Z_B$, of the SM Higgs boson is
\begin{equation}
\Gamma (h \to Z_B Z_B)=\frac{G_B M_h^3 \sin^2 \theta_B}{16 \sqrt{2} \pi} \sqrt{1- 4 x} \left( 1 - 4 x + 12 x^2 \right),
\end{equation}
with $x=M_{Z_B}^2/M_h^2$ and $G_B = 1/ (\sqrt{2} v_B^2)$. 

The three-body decay, $h \to Z_B(p_1) \, q(p_2) \, \bar{q}(p_3)$, is given by
\begin{eqnarray}
\Gamma (h \to Z_B q \bar{q})&=& \frac{1}{(2 \pi)^3} \frac{1}{32 M_h^3} \int_{p_{12}^{\rm min}}^{p_{12}^{\rm max}} d p_{12} \int_{p_{23}^{\rm min}}^{p_{23}^{\rm max}} d p_{23} \  \left| 
\overline{A} ( h \to \bar{q} q Z_B) \right|^2.
\end{eqnarray}
Neglecting the quark masses we have that
\begin{eqnarray}
p_{12}^{\rm min}&=& M_{Z_B}^2, \quad p_{12}^{\rm max}= M_h^2, \\
p_{23}^{\rm min} &=& 0, \hspace{1cm} p_{23}^{\rm max}=\frac{1}{ p_{12}} (p_{12}-M_{Z_B}^2)( M_h^2 - p_{12}), 
\end{eqnarray}
where $p_{ij}=(p_i+p_j)^2$ and the spin-averaged squared amplitude is given by
\begin{eqnarray}
\left| \overline{A} ( h \to \bar{q} q Z_B) \right|^2 &=& \frac{8}{3} \frac{g_B^2}{v_B^2} \frac{M_{Z_B}^4 \sin^2 \theta_B}{\left( (p_{23}- M_{Z_B}^2)^2 + M_{Z_B}^2 \Gamma_{Z_B}^2\right)} \nonumber \\
&\times& \left(   p_{23} + \frac{(p_{12} - M_{Z_B}^2 ) ( M_h^2 - p_{12} - p_{23})}{M_{Z_B}^2}\right).
\end{eqnarray}

\end{itemize}

\section{Production Cross-sections}
\label{app:xsections}

The hadronic production cross-section reads as
\begin{equation}
\sigma (p p \to X Y ) (s) = \int_{\tau_0}^1 d \tau \frac{d {\cal L}^{pp}_{q \bar{q}}}{d \tau}  \ \sigma (q \bar{q} \to XY ) (\hat{s}),
 \end{equation}
 where $\sigma (q \bar{q} \to XY ) (\hat{s})$ corresponds to the partonic cross-section and
\begin{equation}
\label{eq:partonluminosities}
\frac{d {\cal L}^{pp}_{q \bar{q}}}{d \tau} = \int_{\tau}^1 \frac{d x}{x} \left[ f_{q/p} \left(x, \mu \right)  f_{\bar{q}/p} \left(\frac{\tau}{x}, \mu \right) 
+ f_{q/p} \left(\frac{\tau}{x}, \mu \right)  f_{\bar{q}/p} \left(x, \mu \right)\right].
\end{equation}
The parameter $\tau=\hat{s}/s$, where $\hat{s}$ is the partonic center-of-mass energy squared, $s$ is the hadronic center-of-mass energy squared, $\tau_0=(M_X + M_Y)^2/s$ is the production threshold, and $\mu$ is the factorization scale. In what follows we give the analytic results for the partonic cross-sections.

\begin{itemize}
\item Di-quark production: 

\begin{figure}[h]
\centering
\includegraphics[width=0.45\linewidth]{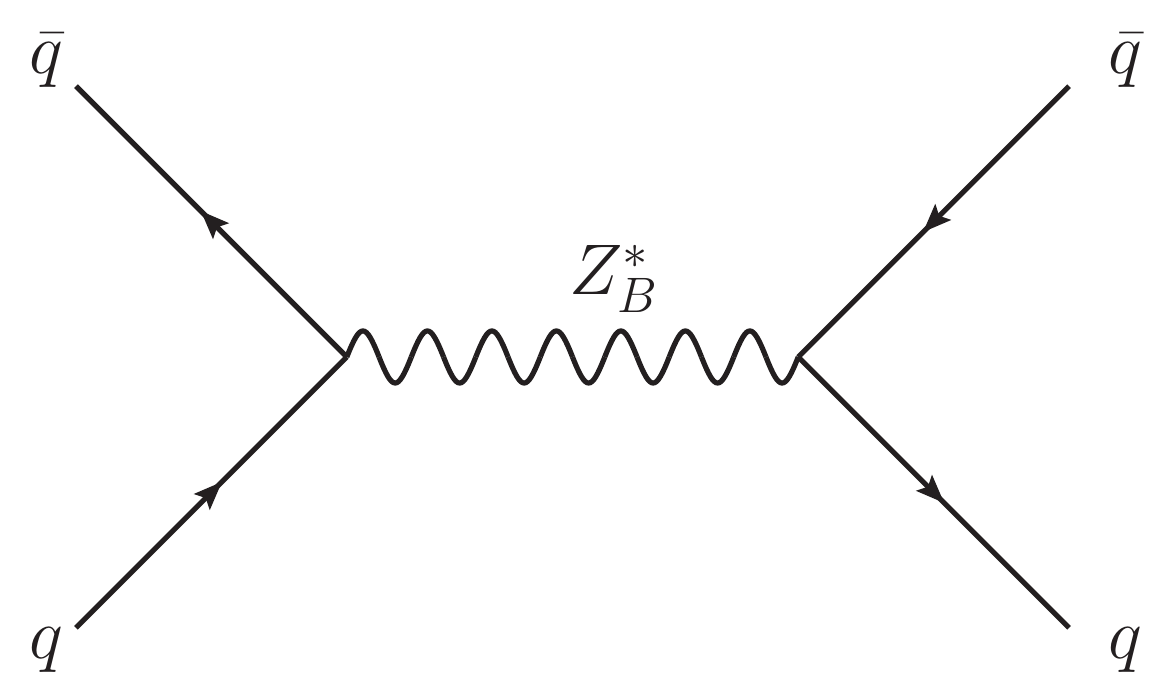}
\caption{Di-quark production channel.}
\end{figure}

The di-quark production cross-section through the leptophobic gauge boson, 
$$ \bar{q}  q \ \to \ Z_B^* \ \to \ \bar{q} q,$$ 
is given by
\begin{equation}
\sigma (\bar{q} q \to Z_B^* \to  \bar{q} q ) (\hat{s})= \frac{g_B^4 \sqrt{\hat{s} - 4 M_q^2}}{972 \pi \sqrt{\hat{s}}} \frac{\left(  2 M_q^2 + \hat{s} \right)}{\left[ (\hat{s} - M_{Z_B}^2)^2 + M_{Z_B}^2 \Gamma_{Z_B}^2 \right]},
\end{equation}
where we have neglected the quark masses in the initial state.

\item Associated Production:

The associated $Z_B - h$ production,
$$pp \  \to \  Z_B^* \  \to \ Z_B  h,$$
is relevant to test the existence of the new Higgs interaction with the leptophobic gauge boson. 

The cross-section at the partonic level is given by
\begin{align}
\sigma (\bar{q} q \to Z_B^* \to  Z_B h ) (\hat{s}) &= \frac{g_B^4  \sin^2 \theta_B}{144 \pi \hat{s}^2} \frac{\left[  \hat{s}^2 - 2 \hat{s} (M_{Z_B}^2 + M_{h}^2) + (M_{Z_B}^2-M_{h}^2)^2 \right]^{1/2}}{\left[ (\hat{s} - M_{Z_B}^2)^2 + M_{Z_B}^2 \Gamma_{Z_B}^2 \right]} \nonumber \\
&\quad \times \left[ \hat{s}^2 + 2 \hat{s} (5 M_{Z_B}^2 - M_{h}^2) + (M_{Z_B}^2-M_{h}^2)^2 \right],
\label{ZBhproduction}
\end{align}
where the $\U(1)_B$ charge of the new scalar is taken as $Q_B\!=\!3$ as in the minimal models~\cite{FileviezPerez:2010gw,FileviezPerez:2011pt,Duerr:2013dza,Perez:2014qfa}.
%
\item Di-boson production:

\begin{figure}[h]
\centering
\includegraphics[width=0.85\linewidth]{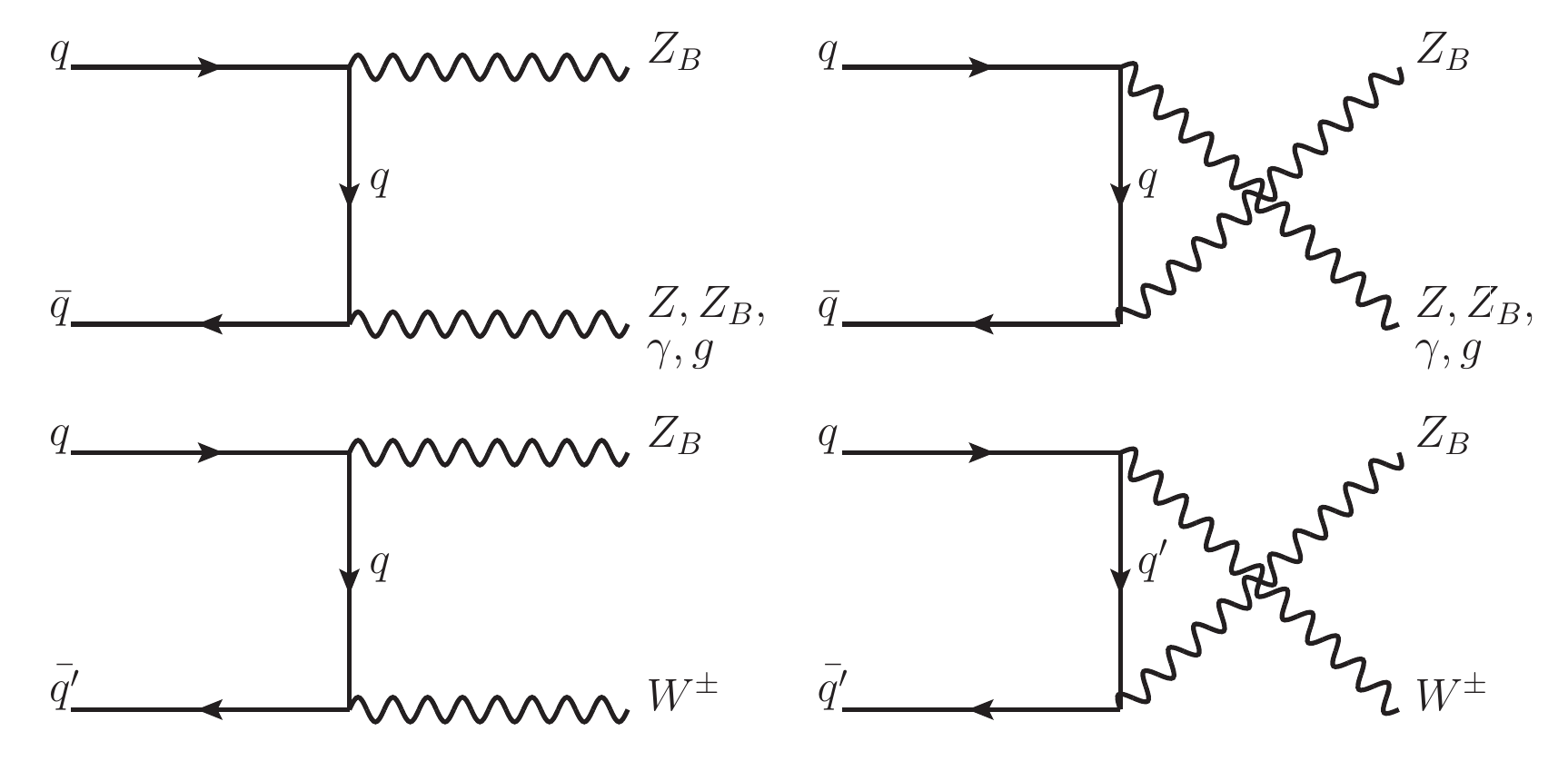}
\caption{Di-boson production channels.}
\end{figure}

Taking the quarks to be massless, the cross-section for the process $q\bar{q}\to Z_B V$ where $V\!=\!Z,\, W^\pm, \, Z_B$ is given by
\begin{align} \label{eq:ZBV}
\sigma(q\bar{q}\to Z_B V)(s) = &  \frac{n \, g_B^2 \left(C_V^2+C_A^2\right) }{ 108 \pi s^2  }  \left[ - 2 \sqrt{f(s)} \right. \nonumber\\[1ex]
& + \left. \frac{ (M_V^2  + M_{Z_B}^2 )^2 + s^2}{s-M_V^2  - M_{Z_B}^2 } \log\left( \frac{  \sqrt{f(s)} + s- M_V^2  - M_{Z_B}^2 } { \sqrt{f(s)} - s+ M_V^2  + M_{Z_B}^2 }\right)   \right]  
\end{align}
where the overall factor $n=1(=1/2$) corresponds to having distinguishable (indistinguishable) particles in the final state,
\beq
f(s) \equiv M_V^4 -2 M_V^2(M_{Z_B}^2+s) + (M_{Z_B}^2-s)^2,
\eeq
and the coefficients $C_V$ and $C_A$ correspond to the vector and axial couplings of the gauge bosons respectively,
\begin{align} \label{GeneralBeta} 
Z_B Z: & \hspace{2mm} C_V  = \frac{g_2}{\cos \theta_W} \left( \frac{1}{2} T_q^3 - Q_q \sin^2 \theta_W \right), \,\,  C_A=-\frac{g_2}{2 \cos \theta_W}T_q^3 \\[2ex]
 Z_B W^\pm: & \hspace{2mm}  C_V = \frac{g_2}{2\sqrt{2}}, \,\,  C_A=-\frac{g_2}{2\sqrt{2}} \\[2ex]
Z_B Z_B  : &\hspace{2mm}  C_V = \frac{g_B}{3}, \,\,  C_A=0.
\end{align}

\item For the process $ q \bar{q} \to Z_B \gamma$ the averaged squared amplitude is given by,
\beq
\big|\overline{\mathcal{M}}(q \bar{q} \to Z_B \gamma)\big|^2 = \frac{2\,e^2\, Q_q^2 \,g_B^2[M_{Z_B}^4-2M_{Z_B}^2t+s^2+2t(s+t)]}{27\, t (M_{Z_B}^2-s-t)},
\eeq
where $Q_q$ corresponds to the electric charge of the quark. In order to compute the proton-proton cross-section we include the cuts on the transverse momentum and the rapidity of the photon (also gluon and quark) as it is explained in the main text.

\item For the process $ q \bar{q} \to Z_B g$ we have 
\beq
\big|\overline{\mathcal{M}}(q \bar{q} \to Z_B g)\big|^2 = \frac{8 \, g_S^2 \, g_B^2[M_{Z_B}^4-2M_{Z_B}^2t+s^2+2t(s+t)]}{81 \, t (M_{Z_B}^2-s-t)},
\eeq
where $g_S$ corresponds to the strong coupling the SM. 

\begin{figure}[h]
\centering
\includegraphics[width=0.85\linewidth]{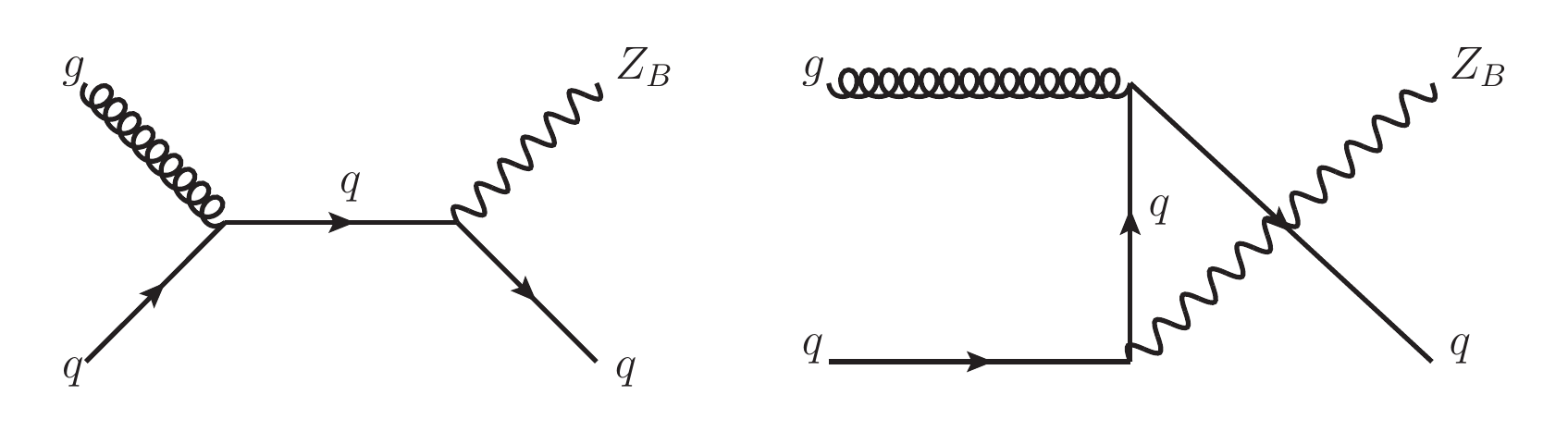}
\caption{Feyman graphs for $ g q \to Z_B q$.}
\end{figure}

\item For the process $q g \to Z_B q$, the averaged squared amplitude is given by
\begin{equation}
\big|\overline{\mathcal{M}}(q g \to Z_B q)\big|^2 =\frac{g_B^2 g_S^2}{27}\frac{M_{Z_B}^4-2M_{Z_B}^2s+2s^2+2st+t^2}{s(s+t-M_{Z_B}^2)},
\end{equation}
and we follow the same procedure as above to compute the proton-proton cross-section. 

\end{itemize}

\newpage
\section{Constraints from Kinetic Mixing}
\label{app:kinmixing}
%
In this Appendix, we study the kinetic mixing between the $\U(1)_Y$ and $\U(1)_B$ gauge bosons, see 
Refs.~\cite{FileviezPerez:2010gw,FileviezPerez:2011pt,Duerr:2013dza,Perez:2014qfa} for realistic theories where 
baryon number is a local gauge symmetry. This parameter can be constrained by studying the properties of the $Z$ boson in the SM, see e.g. \cite{Babu:1997st,Hook:2010tw}. The most general Lagrangian that can be written under the gauge group $\SU(3)_c \otimes \SU(2)_L \otimes \U(1)_Y \otimes \U(1)_B$ involving the neutral gauge bosons of the theory is given by
\begin{align}
\mathcal{L} & \supset  - \frac{1}{4} B_{\mu\nu} B^{\mu\nu} - \frac{1}{2} {\rm{Tr}} \ W_{\mu\nu} W^{\mu\nu} - \frac{1}{4} B'_{\mu\nu} B^{\prime \mu\nu} -\frac{ \sin \epsilon }{2} B_{\mu\nu} B^{\prime\mu\nu}  \nonumber \\[1ex]
 &  + \frac{1}{8} (g_2 W_{3\mu} - g_1 B_\mu) (g_2 W_3^{\mu} - g_1 B^\mu) v_0^2 + \frac{1}{2} \mu_{B'}^2 B'_{\mu} B^{\prime\mu} \label{eq:kinetic} \\[1ex]
 &  -\sum_i \overline{\psi}_i \gamma^{\mu} \left[ g_1 (Y^i_LP_L + Y^i_R P_R)B_{\mu} + g_2 P_L T^a W_{a\mu}\right]\psi_i + g_B \sum_i \overline{\psi}_i \gamma^{\mu} Q_B \psi_i B'_\mu , \nonumber
\end{align}
where $Y_{L/R}$ are the hypercharges of the left/right-handed fields interacting with the hypercharge gauge boson $B_\mu$, $Q_B=1/3$ is the charge of the quarks under the baryon force, $\mu_{B'}=3 g_B v_B$ is the mass term generated after the spontaneously breaking of $\U(1)_B$ and $\sin \epsilon$ parametrizes the kinetic mixing between both Abelian gauge bosons $B_\mu$ and $B_\mu'$.\\

There are different paths to bring the kinetic terms in the first line of Eq.~\eqref{eq:kinetic} to an orthonormal form via a non-orthogonal transformation. For convenience, we choose a change of basis that does not modify the well-known relation between the neutral SM gauge bosons, this can be achieved through the following transformation of the $B_\mu$ and $B_\mu'$ fields:
\begin{equation}
    \begin{pmatrix} B_\mu \\  B'_\mu \end{pmatrix} \mapsto 
    \begin{pmatrix} 1 && \displaystyle -\tan\epsilon \\
                    0 && \displaystyle \sec \epsilon \end{pmatrix}
                    \begin{pmatrix} B_\mu \\ B_\mu' \end{pmatrix}
\end{equation}
which renders the kinetic Lagrangian for the gauge bosons orthonormalized and leads to the following mass terms
\begin{eqnarray}
    {\cal L} &\supset & \frac{1}{8} v_0^2 \left(g_2W_{3\mu} - g_1 (B_\mu - \tan \epsilon B_\mu')\right) \left(g_2 W_{3\mu} - g_1(B^\mu - \tan \epsilon B'^\mu )\right) \nonumber \\ &+& \frac{1}{2}\mu_{B'}^2 \sec^2 \epsilon B_\mu' B'^\mu,
\end{eqnarray}
with the mass matrix in the neutral gauge boson basis $(W^3_\mu, B_\mu, B_\mu')$
\begin{equation}
    M^2_0 = \frac{1}{4} \begin{pmatrix}\displaystyle  g_2^2 v_0^2 && \displaystyle -g_1 \, g_2 v_0^2 && \displaystyle  g_1 \, g_2 \, \tan \epsilon \, v_0^2 \\[2ex]
   - \displaystyle  g_1  \, g_2 \, v_0^2 && \displaystyle   g_1^2 v_0^2 && \displaystyle - g_1^2\tan \epsilon \, v_0^2 \\[2ex]
  \displaystyle g_1 \, g_2 \, \tan \epsilon \, v_0^2 && \hspace{0.25cm} \displaystyle 
   - g_1^2 \tan \epsilon \, v_0^2 && \hspace{0.5cm} \displaystyle  g_1^2 \tan^2\epsilon \, v_0^2 + 4 \mu_{B'}^2\sec^2\epsilon \end{pmatrix}.
\end{equation}

Now, by rotating the $W_3^{\mu}$ and $B^\mu$ fields as it is done in the SM, 
\begin{eqnarray}
\begin{pmatrix} B_\mu \\ \\ W_{3\mu} \end{pmatrix} = \begin{pmatrix}\displaystyle  \cos \theta_W^0 && \displaystyle -\sin  \theta_W^0 \\
&& \\
\displaystyle \sin \theta^0_W && \displaystyle \cos \theta_W^0 \end{pmatrix}\begin{pmatrix} A_\mu \\ \\  C_\mu \end{pmatrix},
\end{eqnarray}
where 
\begin{equation*}
\cos \theta_W^0 \equiv \frac{g_2}{\sqrt{g_1^2+g_2^2}}, \quad \text{and} \quad \sin \theta_W^0 \equiv \frac{g_1}{\sqrt{g_1^2+g_2^2}},
\end{equation*}
the photon decouples and we are left with the following mass matrix for the still unphysical neutral gauge bosons $C_\mu$ and $B_\mu'$:
\begin{equation}
   M^2_0 =  \frac{1}{4} \begin{pmatrix}
    0 & 0 & 0 \\
    & & \\
    0 &\displaystyle  (g_1^2 + g_2^2)v_0^2 & \displaystyle  \sqrt{g_1^2 + g_2^2} \, g_1 \tan \epsilon \, v_0^2 \\
    & & \\
    0 & \hspace{0.3cm} \displaystyle  \sqrt{g_1^2 + g_2^2} \, g_1 \tan \epsilon \, v_0^2  &  \hspace{0.3cm} \displaystyle 4 \mu_{B'}^2 \sec^2 \epsilon + g_1^2 \tan^2 \epsilon \, v_0^2
    \end{pmatrix}.
\end{equation}
\begin{figure}[t]
\centering
\includegraphics[width=0.6\linewidth]{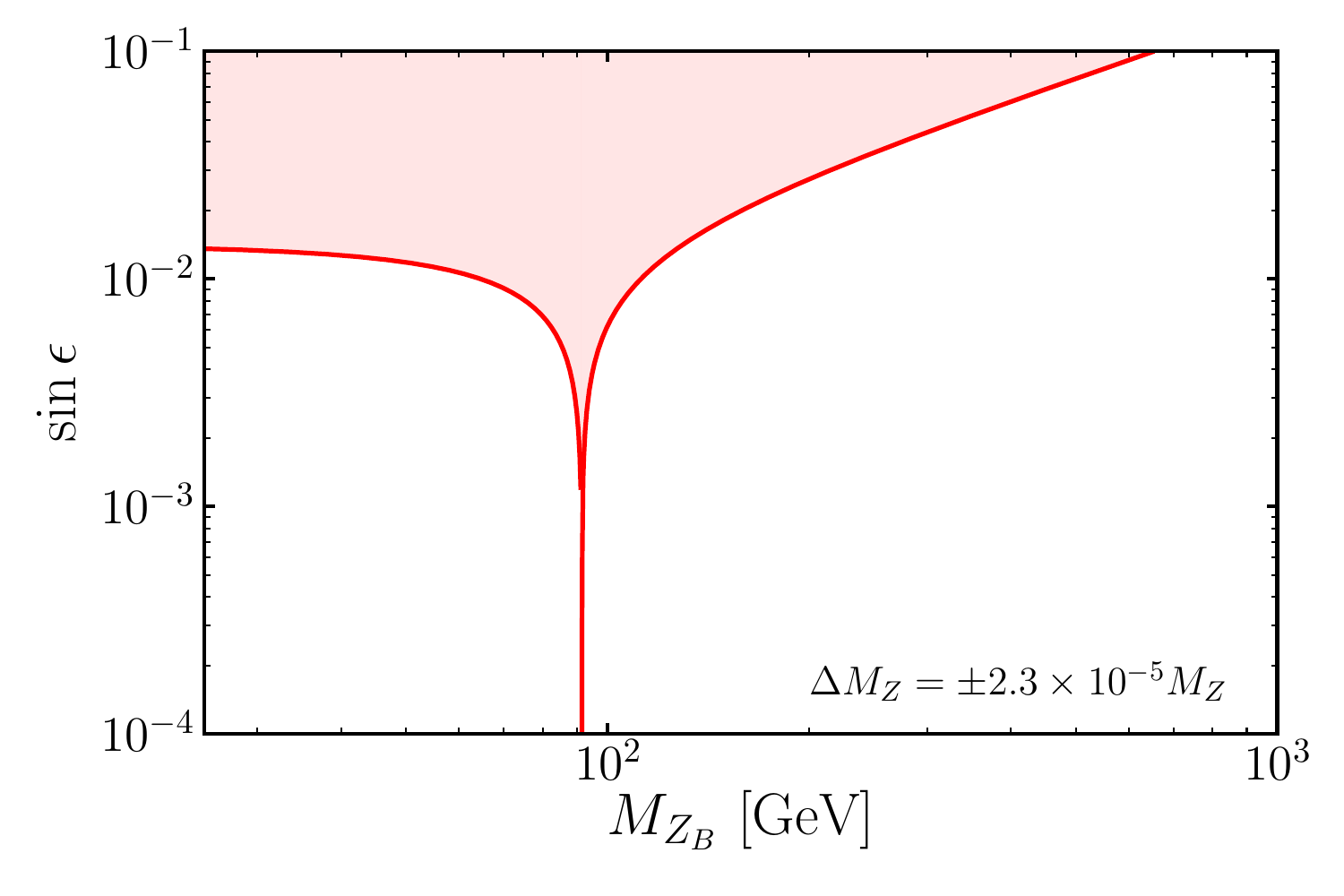}
\caption{ Experimental constraint on the kinetic mixing, $\sin\epsilon$, as a function of the $Z_B$ mass. We have used the measurement of the $Z$ boson mass.}
\label{fig:kinmixing}
\end{figure}

The above mass matrix defines the angle of the final rotation towards the physical basis,
\begin{equation}
\begin{split}
    C_\mu &= \phantom{-} \cos \xi \, Z_\mu + \sin \xi \, Z_{B\mu} \\
    B'_\mu&= -\sin \xi \, Z_\mu + \cos \xi \, Z_{B\mu}
\end{split}
\end{equation}
given by
\begin{equation}
  \text{tan} 2 \xi =\frac{ 2 g_1 \sqrt{g_1^2+g_2^2} \tan \epsilon \, v_0^2 }{  4 \mu_{B'}^2 \sec^2 \epsilon + g_1^2 \tan^2\epsilon \, v_0^2 - (g_1^2 + g_2^2) v_0^2},   
\end{equation}
with the following eigenvalues defining their masses:
\begin{align}
M^2_{A_\mu} = & \, 0, \\
M^2_{Z,Z_B} = & \, \frac{1}{8} \left(g_1^2 \sec^2 \epsilon + g_2^2 \right) v_0^2 + \frac{1}{2} \mu_{B'}^2 \sec^2 \epsilon  \nonumber \\
& \pm \frac{1}{8} \sqrt{ \left(4 \mu_{B'}^2 \sec^2 \epsilon + (g_1^2 \sec^2\epsilon + g_2^2 ) v_0^2 \right)^2 - 16 (g_1^2 + g_2^2) \mu_{B'}^2 v_0^2 \sec^2 \epsilon  },
\end{align}
as expected, in the limit $\epsilon\to 0$ we recover the original masses in the Lagrangian for $Z$ and $Z_B$.

We can now apply the high precision measurement of the $Z$ boson mass to constrain the kinetic mixing parameter, $\sin \epsilon$.
The 1$\sigma$ uncertainty in the experimentally measured $Z$ boson mass is~\cite{Tanabashi:2018oca} 
\beq
\frac{\Delta M_Z}{M^{\rm SM}_{Z}} = \frac{M_Z - M^{\rm SM}_{Z}}{M^{\rm SM}_{Z}} \leq \pm 2.3 \times 10^{-5},
\eeq
and this can be used to constrain the shift induced by the kinetic mixing. 
In Fig.~\ref{fig:kinmixing} we show this constraint in the $M_{Z_B}$ vs $\sin \epsilon$ plane; as can be seen, the kinetic mixing has to be very small and it does not change the main results in our paper. Recently, the CMS~\cite{CMS:2019kiy} and the LHCb~\cite{Aaij:2017rft} collaborations found stronger constraints for this mixing parameter for $M_{Z_B}\leq 200$ GeV by searching for the direct production of a new gauge boson.

\bibliographystyle{JHEP}
\bibliography{Higgs-ZB}{}
\end{document}